\documentclass[aps,showpacs,amsmath,amssymbd,dvips,showkeys,linenumbers,preprint,preprintnumbers]{revtex4}
\usepackage{graphicx}
\usepackage{color}
\usepackage{fancyhdr}
\def \be  {\begin{equation}}
\def \ee  {\end{equation}}
\def \ee  {\end{equation}}
\def \bea {\begin{eqnarray}}
\def \eea {\end{eqnarray}}

\newcommand{\nn}{\nonumber}

\begin{document}

\preprint{ECTP-2016-04}
\preprint{WLCAPP-2016-04}
\hspace*{3mm}

\title{SU($3$) Polyakov linear-sigma model: Conductivity and viscous properties of QCD matter in thermal medium}
\author{Abdel Nasser Tawfik}
\email{atawfik@rcf.rhic.bnl.gov}
\affiliation{Egyptian Center for Theoretical Physics (ECTP), Modern University for Technology and Information (MTI), 11571 Cairo, Egypt}
\affiliation{World Laboratory for Cosmology And Particle Physics (WLCAPP), 11571 Cairo, Egypt}
\author{Abdel Magied Diab} 
\affiliation{Egyptian Center for Theoretical Physics (ECTP), Modern University for Technology and Information (MTI), 11571 Cairo, Egypt}
\affiliation{World Laboratory for Cosmology And Particle Physics (WLCAPP), 11571 Cairo, Egypt}

\author{M. T. Hussein}
\affiliation{Physics Department, Faculty of Science,  Cairo University, 12613 Giza, Egypt}

\begin{abstract}

In mean field approximation, the grand canonical potential  of SU(3) Polyakov linear-$\sigma$ model (PLSM) is analysed for chiral phase-transition, $\sigma_l$ and $\sigma_s$ and for deconfinement order-parameters, $\phi$ and $\phi^*$ of light- and strange-quarks, respectively. Various PLSM parameters are determined from the assumption of global minimization of the real part of the potential. Then, we have calculated the subtracted condensates ($\Delta_{l,s}$). All these results are compared with recent lattice QCD simulations. Accordingly, essential PLSM parameters are determined. The modelling of the relaxation time is utilized in estimating the conductivity properties of the QCD matter in thermal medium, namely electric [$\sigma_{el}(T)$] and heat [$\kappa(T)$] conductivities. We found that the PLSM results on the electric conductivity and on the specific heat agree well with the available lattice QCD calculations. Also, we have calculated bulk and shear viscosities normalized to the thermal entropy, $\xi/s$ and $\eta/s$, respectively, and compared them with recent lattice QCD. Predictions for $(\xi /s)/(\sigma_{el}/T)$ and  $(\eta/s)/(\sigma_{el}/ T)$ are introduced.  We conclude that our results on various transport properties show some essential ingredients, that these properties likely come up with, in studying QCD matter in thermal and dense medium. 

\end{abstract}

\pacs{12.39.Fe, 25.75.Nq, 12.38.Mh, 67.10.Jn}
\keywords{Chiral Lagrangian, Quark confinement, Quark-gluon plasma, Transport processes in quantum fluid}

\maketitle
\tableofcontents
\makeatletter
\let\toc@pre\relax
\let\toc@post\relax                 
\makeatother 

\section{Introduction \label{Intro}}

The characterization of the electro-magnetic properties of hadron and parton matter, which in turn can be described by quantum chromodynamics (QCD) and quantum electrodynamics (QED), gains increasing popularity among particle physicists. One of the main gaols of the relativistic heavy-ion facilities such as the Relativistic Heavy-Ion Collider (RHIC) at BNL, Uppton-USA and the Large Hadron Collider (LHC) at CERN, near Geneva-Switzerland and the future Nuclotron-based Ion Collider fAcility (NICA) at JINR, Dubna-Russia, is precise determination of the hadron-parton phase-diagram, which can also be studied in lattice QCD numerical simulations \cite{Gross:1973, Politzer:1973} and various QCD-like approaches. The Polyakov Nambu-Jona Lasinio (PNJL) model \cite{Fukushima:2004, Fukushima:2008, Ratti:2005}, the Polyakov linear-$\sigma$ model (PLSM) or the Polyakov quark meson model (PQM) \cite{Schaefer:2007c, Schaefer:2007d, Schaefer:2008ax, Schaefer:2009ab}, and the Dynamical Quasi-Particle model (DQPM) \cite{Schneider:2004,Letessier:2003,Bluhm:2005} are examples on QCD-like models aiming to characterizing the strongly interacting matter in dense and thermal medium and also in finite electro-magnetic field. 

It is conjectured that, the [electrical and thermal (heat)] conductivity and (bulk and shear) viscous properties of the QCD matter come up with significant modifications in the chiral phase-transition \cite{Preis:2011, Bruckmann2013i, Huang2012p}. The influence of finite magnetic field on QCD phase-diagram, which describes the variation of the confinement-deconfinement phase-transition at various baryon chemical potentials \cite{Fodor:nature2006}, has been studied in lattice QCD \cite{Fodor2012pio}. In relativistic heavy-ion collisions, a huge magnetic field can be created due to the relativistic motion of charged spectators and the local momentum-imbalance of the participants. At LHC energy, the expected magnetic field $\sim 10 m_{\pi}^2$ \cite{Skokov:2009, Elec:MagnetA,Elec:MagnetB}, where $m_{\pi}^2 \sim 10^8$ Gauss. In order to estimate the temperature dependence of the electrical conductivity, different phenomenological approaches have been proposed \cite{Fernandez:2006u, Steinert:2014sa, Cassing:2014sai}. Besides electrical conductivity, the magnetic catalysis, for instance, is found sensitive to the response of the strongly interacting system to finite electro-magnetic field \cite{Tawfik:2014hwa,Ezzelarab:2015tya,Tawfik:2015apa,Tawfik:2016lih,Tawfik:2016gye,Tawfik:2016cot, TDthermo,THD:GEM2015}. 

The chiral phase-structure of various mesonic states at finite temperatures has been evaluated with and without anomaly contributions \cite{Schaefer:2008hk,TiwariA,TiwariB} and in presence of finite magnetic fields \cite{THD:GEM2015}. In a previous work, we have presented calculations for the chiral phase-structure of (pseudo)-scalar and (axial)-vector meson masses in thermal and dense medium with and without Polyakov corrections and/or anomaly contributions \cite{Tawfik:2014gga} a vanishing and finite magnetic effect \cite{THD:GEM2015}. The chiral phase-structure in the limit of large number of colors ($N_c$) and the normalization of sixteen meson states with respect to the lowest Matsubara frequency are introduced in Ref. \cite{Tawfik:2014gga}. In finite magnetic field, the chiral phase-structure of (pseudo)-scalar and (axial)-vector meson masses has been analysed \cite{THD:GEM2015}. Recently, study of QGP in presence of external magnetic field has been conducted \cite{Tawfik:Magnetic,THD:GEM2015, TSHfiniteB}. Furthermore, at nonzero magnetic field, viscous properties from Boltzmann-Uehling-Uhlenbeck (BUU) equation have been compare with the ones from Green-Kubo (GK) correlations in relaxation time approximation (RTA), which are based on relativistic kinetic theory \cite{TSHfiniteB}. 

Some QCD transport coefficients have been determined, numerically and estimated, analytically \cite{Rischke1, Rischke2, Rischke3}. The confrontation to lattice QCD results enables the judgement about the QCD-effective models, such as PNJL and PLSM. The transport coefficients calculated from PNJL \cite{Bratkovskaya} and DQPM \cite{Bratkovskaya:2014} and thermodynamics and bulk viscosity near phase transition from $\mathcal{Z}(1)$ and $\mathcal{O}(4)$ models in Hartree approximation for Cornwall-Jackiw-Tomboulis (CJT) formalism are summarized in Ref. \cite{Huang:2009}. The calculations of shear and bulk viscosities of  hadrons \cite{Tawfik:2010mb} and that of both hadrons and partons from parton-hadron string dynamics (PHSD) were presented in Ref. \cite{Bratkovskaya:2013}. The ratios of bulk and shear viscosity to the electrical conductivity of QGP were determined \cite{Puglisi:2014}. 

The transport coefficients are particularly helpful in characterizing QCD matter, such as the phase transition, the critical endpoint, etc. \cite{Kapusta:1993A,Kapusta:1993A}. Recent non-perturbative lattice QCD simulations succeeded in estimating QCD viscosities. We examine the [electrical and thermal (heat)] conductivities and (bulk and shear) viscosities as diagnostic tools to studying quark-hadron phase-transition in thermal medium. The viscous properties have been reported in Ref. \cite{Kapusta:2008}.  We recall that the so-far different LSM-calculations have been performed in order to determine certain transport-properties of the QCD matter \cite{Dobado:2009, Dobado:2012, Chakraborty:2011}.  While the system approaches equilibrium, the temperature dependence of the relaxation time has been characterized. In light of this, studying the QCD regimes, where analytic calculations can be compared with, is of great relevance to recent problems in high-energy physics. This would clarify the validity of the proposed calculations, in this case QCD-like approaches such as PLSM,  in determining other quantities in dense medium and measure the influence of finite electro-magnetic field. 

Before introducing the present results, the question to what extent the transport coefficients are sensitive to the underlying microscopic physics of the medium? should be answered, first. Its answer determines how relevant is the present work in describing recent lattice QCD simulations. Both lattice QCD calculations and ours from the QCD-like approach, PLSM, share almost same approximations, for instance, both assume a global "equilibrium". In other words, even if nowadays the first-principle lattice QCD calculations become very reliable, and they are not "dynamical" at all. The lattices are static assuming nonvarying temporal and spacial dimensions. Anisotropic lattices are technically not yet possible. Thus, the calculations of certain transport coefficients on lattices explicitly assumes that the QCD system is in equilibrium. So, do our QCD-like approach, the PLSM, which integrates some features and symmetries of QCD and the corresponding degrees-of-freedom. In light of this, the answer to the above mentioned question becomes simple. Yes, we can extend "static approaches'' utilized in calculating thermodynamics, for instance, to ''transport properties'', as long as we assume equilibrium. The role of an effective translator allowing us to calculate transport coefficients from PLSM is played by the relaxation time. A good modelling for the relaxation time is therefore very crucial in enabling PLSM to reproduce lattice QCD transport coefficients.

In the present work, we introduce SU($3$) PLSM calculations for electrical and heat conductivities and for bulk and shear viscosities at finite temperature but vanishing baryon chemical potential. To this end, we determine various thermodynamic quantities, such as, the equation of state, specific heat, squared speed of sound and quark number multiplicity as functions of temperatures \cite{TDthermo}.  The PLSM approach shall be elaborated in section \ref{sec:model}. Section \ref{structure} summarizes the chiral phase-structure from PLSM and compares the phase transition for light and strange quarks with recent lattice QCD calculations. The relaxation time shall be calculated in section \ref{sec:relxtime}. The normalized electrical and heat conductivities are presented in section \ref{conductivitya}. The ratios of bulk and shear viscosities relative to the thermal entropy shall be given in section \ref{Viscosity}.  Section \ref{Conclusion} is devoted to the conclusions.

\section{A short reminder to SU($3$) Polyakov linear-sigma model \label{sec:model}}  

The PLSM Lagrangian with $N_f =3$ quark flavors and $N_c =3$ color degrees-of-freedom consists of two parts
\begin{eqnarray}
\mathcal{L}=\mathcal{L}_{chiral}-\mathbf{\mathcal{U}}(\phi, \phi^*, T). \label{plsm}
\end{eqnarray}
The chiral part $\mathcal{L}_{chiral}=\mathcal{L}_q+\mathcal{L}_m$, which is coupled to the Polyakov-loop potential with SU(3)$_{L} \times$ SU(3)$_{R}$ symmetry  \cite{Lenaghan,Schaefer:2008hk}, in turn consists of  the fermionic contributions from quarks, $\mathcal{L}_f$, coupled with a flavor-blind Yukawa coupling $g$ of the quarks \cite{blind} and of the mesonic contributions from gluons. 
\begin{itemize}
\item The fermionic part reads
\bea
\mathcal{L}_q &=& \sum_f \overline{q}_f(i\gamma^{\mu} D_{\mu}-gT_a(\sigma_a+i \gamma_5 \pi_a)) q_f, \label{lfermion}
\eea
where $\gamma^{\mu}$ are the Dirac gamma matrices of  chiral spinors, $\sigma_a$ are the scalar mesons and $\pi_a$ are the pseudoscalar mesons. Through the covariant derivative $D_{\mu}=\partial_{\mu}-i A_{\mu}$, the quarks can be coupled to the Euclidean gauge field \cite{Polyakov:1978vu, Susskind:1979up} $A_{\mu}=\delta_{\mu 0} A_0$. 

\item The second term, $\mathcal{L}_m$, refers to the mesonic contributions,
\bea
\mathcal{L}_m &=& \mathrm{Tr}(\partial_{\mu}\Phi^{\dag}\partial^{\mu}\Phi-m^2
\Phi^{\dag} \Phi)-\lambda_1 [\mathrm{Tr}(\Phi^{\dag} \Phi)]^2 
-\lambda_2 \mathrm{Tr}(\Phi^{\dag}
\Phi)^2 \nn \\
&+& c[\mathrm{Det}(\Phi)+\mathrm{Det}(\Phi^{\dag})]
+\mathrm{Tr}[H(\Phi+\Phi^{\dag})],  \label{lmeson}
\eea
with $\Phi$ is a complex $3\times3$ matrix and defined as \cite{Schaefer:2008hk}, 
\bea
\Phi =T_a(\sigma_a+i\pi_a),
\eea
where $T_a=\lambda_a/2$ with $a = 0, \cdots, 8$ are nine generators of $U(3)$ symmetry group. $\lambda_a$ are Gell-Mann matrices while $\lambda_0 = \sqrt{2/3} \,{\bf \hat{I}}$ \cite{Gell Mann:1960}. 
Explicit symmetry-breaking is given by $H = T_a h_a$, which is ($3 \times 3$) matrix with nine parameters $h_a$. 
\end{itemize}
In vacuum, the model parameters can be fixed  by six experimentally known quantities. Tab. \ref{tab:1a} summarizes estimations for these parameters at sigma mass $m_\sigma =800~$MeV \cite{Schaefer:2008hk}.

\begin{table}[htb]
\begin{center}
\begin{tabular}{|c | c | c | c | c | c | c |}
\hline
$m_\sigma$ [MeV] & $c\,$ [MeV] & $h_l\,$ [MeV$^3$] & $h_s\,$ [MeV$^3$] & $m^2 \,$ [MeV$^2$] & $\lambda _1$ & $\lambda _2$\\ 
\hline 
800 & $4807.84$ & $(120.73)^3$ & $(336.41)^3$ & -$(306.26)^2$ & $13.49$& $46.48$\\ 
\hline 
\end{tabular}
\caption{Summary of PLSM's parameters. A detailed description is given in Ref. \cite{Schaefer:2008hk}  \label{tab:1a}}
\end{center}
\end{table} 


Through integrating the Polyakov-loop variables $\phi$ and $\phi^*$ for light and strange quarks, respectively, the Polyakov-loop potential introduces gluonic degrees-of-freedom and the dynamics of the quark-gluon interactions to the QCD-like matter. Various expressions fulfilling QCD symmetries in pure-gauge theory have been proposed \cite{Ratti:2005, Roessner:2007, Schaefer:2007d, Fukushima:2008}. In all our previous works \cite{Tawfik:LSM, Tawfik:2014gga, Tawfik:Magnetic,Tawfik:quasi}, we have utilized the simplest polynomial-form. But, in the present work, we introduce results based in the alternatively-{\it improved} extension \cite{Roessner:2007, Sasaki:2013ssdw}; the logarithmic potential, 
\bea
\frac{\mathbf{\mathcal{U}}_{\mathrm{Log}}(\phi, \phi^*, T)}{T^4} = \frac{-a(T)}{2} \; \phi^* \phi + b(T)\; \ln{\left[1- 6\, \phi^* \phi + 4 \,( \phi^{*3} + \phi^3) - 3 \,( \phi^* \phi)^2 \right]}, \label{LogULoop}
\eea 
with 
\bea
a(T) = a_0 + a_1 \left(T0/T\right) + a_2 \left(T0/T\right)^2  \qquad  \mathrm{and}  \qquad b(T) =
 b_3  \left(T0/T\right)^3,
\eea
where $T_0$ is the critical temperature for the deconfinement phase-transition in the pure-gauge sector.  This potential is qualitatively consistent with the leading-order results from strong coupling expansion \cite{Langelage2011}.  Furthermore,  the fact that  at high temperature, $\phi$ and $\phi^* \rightarrow 1$ and Eq. (\ref{LogULoop}) approaches a maximum value sets limits to the Polyakov-loop variables. Accordingly, one enforces the Polyakov loop in the target region respecting the SU($3$) structure of the theory. The parameters $a_0$, $a_1$, $a_2$, and $b_3$ can be determined through reproduction of lattice pure-gauge thermodynamics. The results are listed  in Tab. \ref{FitparameterPLOY}.  

The authors of  Ref. \cite{Sasaki:2013ssdw} have reestimated the fit parameters to recent lattice QCD calculations. In their polynomial-logarithmic parametrisation of the Polyakov-loop potential, they even included higher-order terms, 
\bea
\frac{\mathbf{\mathcal{U}}_{\mathrm{PolyLog}}(\phi, \phi^*, T)}{T^4} &=&   \frac{-a(T)}{2} \; \phi^* \phi + b(T)\; \ln{\left[1- 6\, \phi^* \phi + 4 \,( \phi^{*3} + \phi^3) - 3 \,( \phi^* \phi)^2 \right]} \nn \\ &+& \frac{c(T)}{2}\, (\phi^{*3} + \phi^3) + d(T)\, ( \phi^* \phi)^2. \label{LogPloy}
\eea 
It is obvious that if $c(T)$ and $d(T)$ vanish, Eq. (\ref{LogPloy}) reduces to Eq. (\ref{LogULoop}).  The various coefficients in Eq. (\ref{LogPloy}) have been determined \cite{Sasaki:2013ssdw},
\bea
x(T) = \frac{x_0 + x_1 \left(T0/T\right) + x_2 \left(T0/T\right)^2}{1+x_3 \left(T0/T\right) + x_4 \left(T0/T\right)^2},  \qquad  \qquad  b(T)= b_0\, \left(T0/T\right)^{b_1} \left(1-e^{b_2 \left(T0/T\right)^{b_3}} \right),
\eea 
where $x=(a,\,c,\,d)$. The different parameters  are summarised in Tab. \ref{FitparameterPLOY}. Equation (\ref{LogPloy}) takes into account the fluctuations of the Polyakov loop in addition to the expectation value of the Polyakov loop and the pressure. Furthermore, if the fluctuations of the gluonic observables are involved, both Eq. (\ref{LogULoop}) and Eq. (\ref{LogPloy}) give different results.

\begin{table}[htb]
\begin{center}
 \begin{tabular}{p{2cm} p{2cm}  p{2cm} p{2cm} p{2cm} p{1.5cm}} 
 \hline \hline
 Ref. \cite{Roessner:2007} & $a_0$ & $a_1$ & $a_2$ & $b_3$ \\ [0.5ex] 
 & $3.51$ & $-2.47$ &  $15.2$ & $-1.75$ \\
 \hline \hline
 Ref. \cite{Sasaki:2013ssdw} & $a_0$ & $a_1$ & $a_2$ & $a_3$ & $a_4$ \\
 & $-44.14$ & $151.4$ & $-90.0677$ & $2.77173$ & $3.56403$ \\
 \hline 
 & $b_0$ & $b_1$ & $b_2$ & $b_3$ & \\
 & $-0.32665$ & $5.85559$ & $-82.9823$ & $3.0$ & \\
 \hline 
 & $c_0$ & $c_1$ & $c_2$ & $c_3$ & $c_4$ \\
 & $-50.7961$ & $114.038$ & $-89.4596$ & $3.08718$ & $6.72812$ \\
 \hline 
 & $d_0$ & $d_1$ & $d_2$ & $d_3$ & $d_4$ \\
 & $27.0885$ & $-56.0859$ & $71.2225$ & $2.9715$ & $6.61433$ \\
 \hline  \hline
\end{tabular}
\caption{Fit parameters deduced from logarithmic \cite{Roessner:2007} and polynomial-logarithmic Polyakov-loop potentials \cite{Sasaki:2013ssdw} with recent lattice QCD simulations. \label{FitparameterPLOY}}
\end{center} 
\end{table}


In thermal equilibrium, the exchanges of energy between particles and antiparticles in  PLSM can be described by path integral over quark, antiquark and meson fields. The well-known pure gauge results (not shown here) can be produced. Thus, the grand canonical partition function $\mathcal{Z}$ reads   
\begin{eqnarray}
\mathcal{Z}&=& \mathrm{Tr \,exp}[-(\hat{\mathcal{H}}-\sum_{f=u, d, s} \mu_f \hat{\mathcal{N}}_f)/T] \nonumber\\
&=& \int\prod_a \mathcal{D} \sigma_a \mathcal{D} \pi_a \int
\mathcal{D}\psi \mathcal{D} \bar{\psi} \mathrm{exp} \left[ \int_x
(\mathcal{L}+\sum_{f=u, d, s} \mu_f \bar{\psi}_f \gamma^0 \psi_f )
\right],
\end{eqnarray}
where $\int_x\equiv i \int^{1/T}_0 dt \int_V d^3x$ with $V$ being the volume and $\mu_f$ the chemical potential for quark flavors $f=(u, d, s)$. We assume symmetric quark matter and degenerate light quarks and therefore define a uniform blind chemical potential $\mu_f \equiv \mu_{u, d} = \mu_s$ \cite{Schaefer:2007c, Schaefer:2008hk, blind}. 

For meson fields,  we can implement the expectation values $\bar{\sigma_l}$ and $\bar{\sigma_s}$, for light and strange quark, respectively \cite{Kapusta:2006pm,Mao:2010}. Standard methods are used in calculating the integrals over the fermions yields \cite{Kapusta:2006pm}. Then, the thermodynamic potential density $ \Omega (T, \mu)=-T \cdot \ln{\left[\mathcal{Z}\right]}/V$ or 
\begin{eqnarray}
\Omega(T, \mu) 
&=& U(\sigma_l, \sigma_s)+\mathbf{\mathcal{U}}(\phi, \phi^*, T)+\Omega_{\bar{q}q}. \label{potential}
\end{eqnarray}
The purely mesonic potential is given as
 \begin{eqnarray}
U(\sigma_l, \sigma_s) &=& - h_l \sigma_l - h_s \sigma_s + \frac{m^2\, (\sigma^2_l+\sigma^2_s)}{2} - \frac{c\, \sigma^2_l \sigma_s}{2\sqrt{2}}  
+ \frac{\lambda_1\, \sigma^2_l \sigma^2_s}{2} +\frac{(2 \lambda_1
+\lambda_2)\sigma^4_l }{8}+ \frac{(\lambda_1+\lambda_2)\sigma^4_s}{4}. \hspace*{8mm} \label{Upotio}
\end{eqnarray}

In mean field approximation, the thermodynamic potential for the quarks and antiquarks contributions was introduced by Fukushima \cite{Fukushima:2008} and other authors \cite{Kapusta:2006pm, Mao:2010, Schaefer:2009sasd},
\bea
\Omega_{\bar{q}q}(T, \mu _f)&=& -2 \,T \sum_{f=l, s} \int_0^{\infty} \frac{d^3\vec{P}}{(2 \pi)^3} \left\{ \ln \left[ 1+3\left(\phi^*+\phi\, e^{-\frac{E_f-\mu _f}{T}}\right)\, e^{-\frac{E_f-\mu _f}{T}}+e^{-3 \frac{E_f-\mu _f}{T}}\right] \right. \nonumber \\ 
&& \hspace*{35mm} \left.  +\ln \left[ 1+3\left(\phi+\phi^*\, e^{-\frac{E_f+\mu _f}{T}}\right)\, e^{-\frac{E_f+\mu _f}{T}}+e^{-3 \frac{E_f+\mu _f}{T}}\right] \right\}, \hspace*{8mm} \label{PloykovPLSM}
\eea
where $E_f=(\vec{P}^2+m_f^2)^{1/2}$ is the dispersion relation corresponding to quark and antiquark and $m_f$ is the flavor mass of light and strange quark, where $m_l = g \sigma_l/2$ and $m_s = g \sigma_s/\sqrt{2}$  \cite{Kovacs:2006}. The subscripts $l$ and $s$ refer to degenerate light and strange quark, respectively.

Eqs. (\ref{LogPloy}), (\ref{Upotio}) and (\ref{PloykovPLSM})  construct the thermodynamic potential density, Eq. (\ref{potential}), in which seven parameters $m^2, h_l, h_s, \lambda_1, \lambda_2, c$ and $ g$,  condensates $\sigma_l$ and $ \sigma_s$ and an order parameters for deconfinement $\phi$ and $\phi^*$ should be determined. First, the six parameters $m^2, h_l, h_s, \lambda_1, \lambda_2 $ and $c$ can be fixed in vacuum by six experimentally well-known quantities \cite{Schaefer:2008hk}. 

In order to evaluate  the  expectation values of the PLSM order-parameters, $\sigma_l=\bar{\sigma_l}$, $ \sigma_s=\bar{\sigma_s}$, $\phi=\bar{\phi}$ and $\phi^*=\bar{\phi^*}$, one can minimize the thermodynamic potential in Eq. (\ref{potential}) as a follow.
\begin{eqnarray}
\left.\frac{\partial \Omega}{\partial \bar{\sigma_l}}= \frac{\partial
\Omega}{\partial \bar{\sigma_s}}= \frac{\partial \Omega}{\partial
\bar{\phi}}= \frac{\partial \Omega}{\partial \bar{\phi^*}}\right|_{min} =0. \label{cond1}
\end{eqnarray}
However, the PLSM thermodynamic potential,  Eq. (\ref{potential}), is complex in nonzero chemical-potential ($\mu\ne0$) and in finite Ployakov-loop variables. A minimization of a complex function would be seen as void of meaning. An analysis of the order parameters is given by minimizing the real part of thermodynamic potential ($\mbox{Re}\; \Omega$). In principle, the (thermal) expectation values of Ployakov-loop $\bar{\phi}$ and its conjugate $\bar{\phi}^*$ must be real quantities as discussed in Ref. \cite{Dumitru2005ss}. The solutions of these equations can be determined by minimizing the real potential at a saddle point. They determine the behavior of the chiral order-parameter $\bar{\sigma_l}$, $\bar{\sigma_s}$ and the Polyakov-loop expectation values $\bar{\phi},\; \bar{\phi}^*$ as functions of $T$ and $\mu$. 

It should be noticed that we have solved the expressions (\ref{cond1})  as a complete set of equations. Alternatively, their individual solutions (not shown here), which become relevant at finite $\mu$ where the gap equations are well-defined, can be derived. Both solutions are almost identical.

\section{Results}

In this sections, we introduce the SU($3$) PLSM calculations for  order-parameters [section \ref{structure}], relaxation time [section \ref{sec:relxtime}], electrical and heat conductivity [section \ref{conductivitya}], and bulk and shear viscosity [section \ref{Viscosity}] of QCD matter in thermal medium. A vanishing baryon chemical potential is assumed in all these calculations.

\subsection{Phase transition(s) and relaxation time \label{results}} 

In order to calculate the conductivity and viscous properties from PLSM, other quantities including order parameters and relaxation time or decay constant are needed. In the section that follows, we estimate from PLSM the chiral and deconfinement phase-transitions, i.e. the PLSM order parameters. Then, we introduce the temperature dependence of the relaxation time at different values of the chemical potentials. Then, we determine the transport properties such as electrical and heat conductivity and the viscous properties. The results are  confronted to recent lattice QCD simulations whenever available.

\subsubsection{Chiral and deconfinement order-parameters \label{structure}}

\begin{figure}[htb] 
\includegraphics[width=3.8cm,angle=-90]{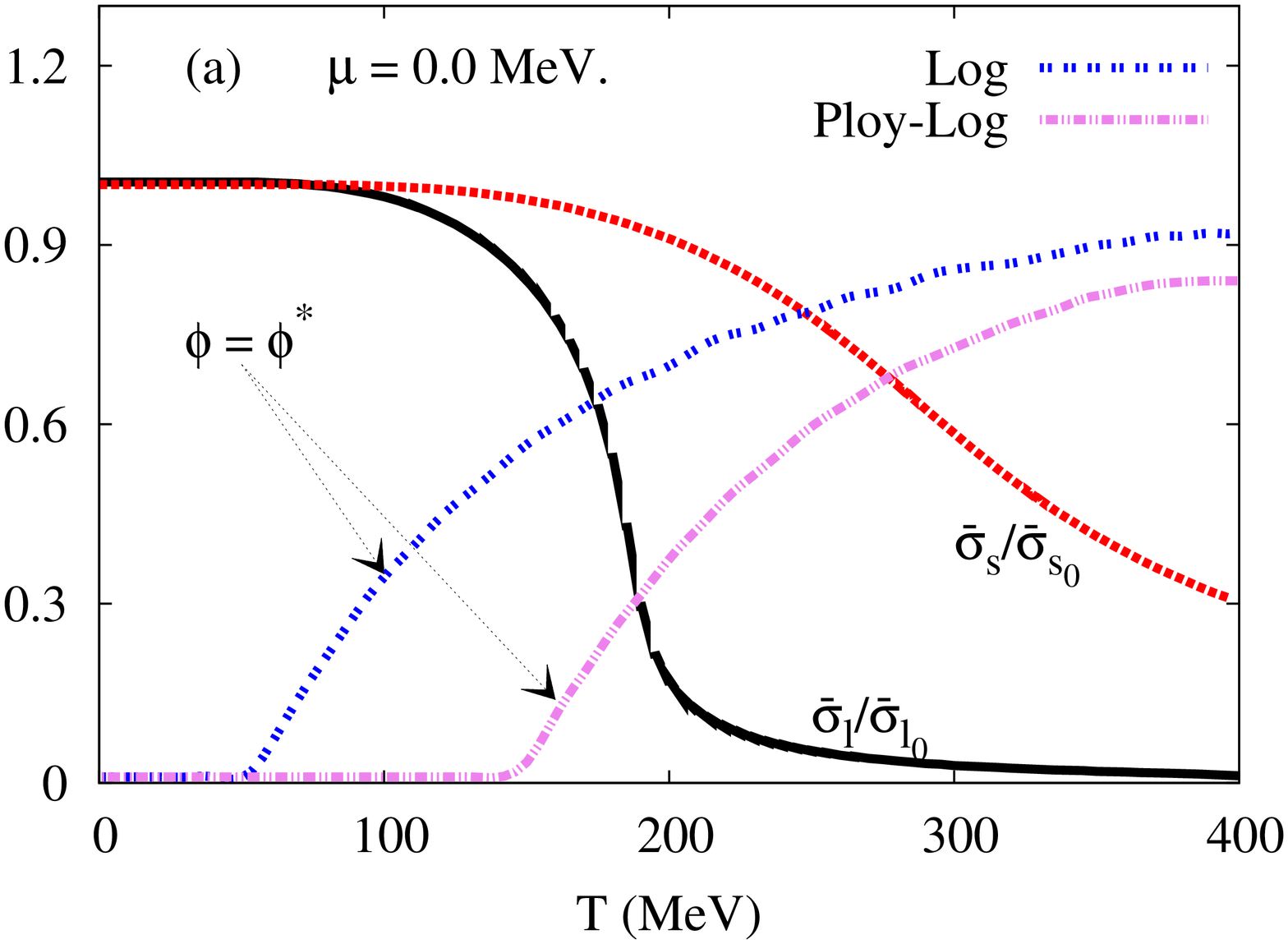}
\includegraphics[width=3.8cm,angle=-90]{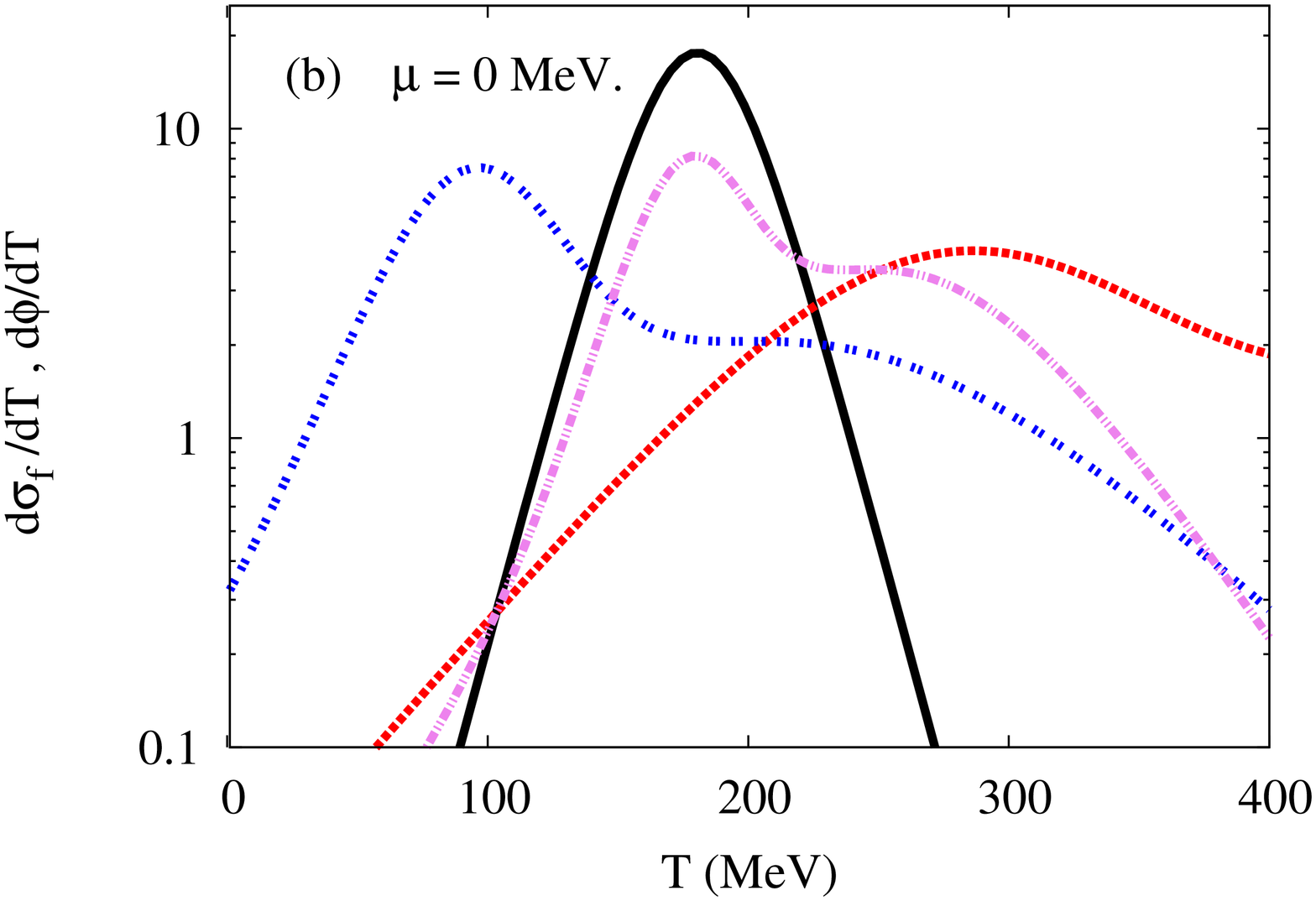}
\includegraphics[width=3.8cm,angle=-90]{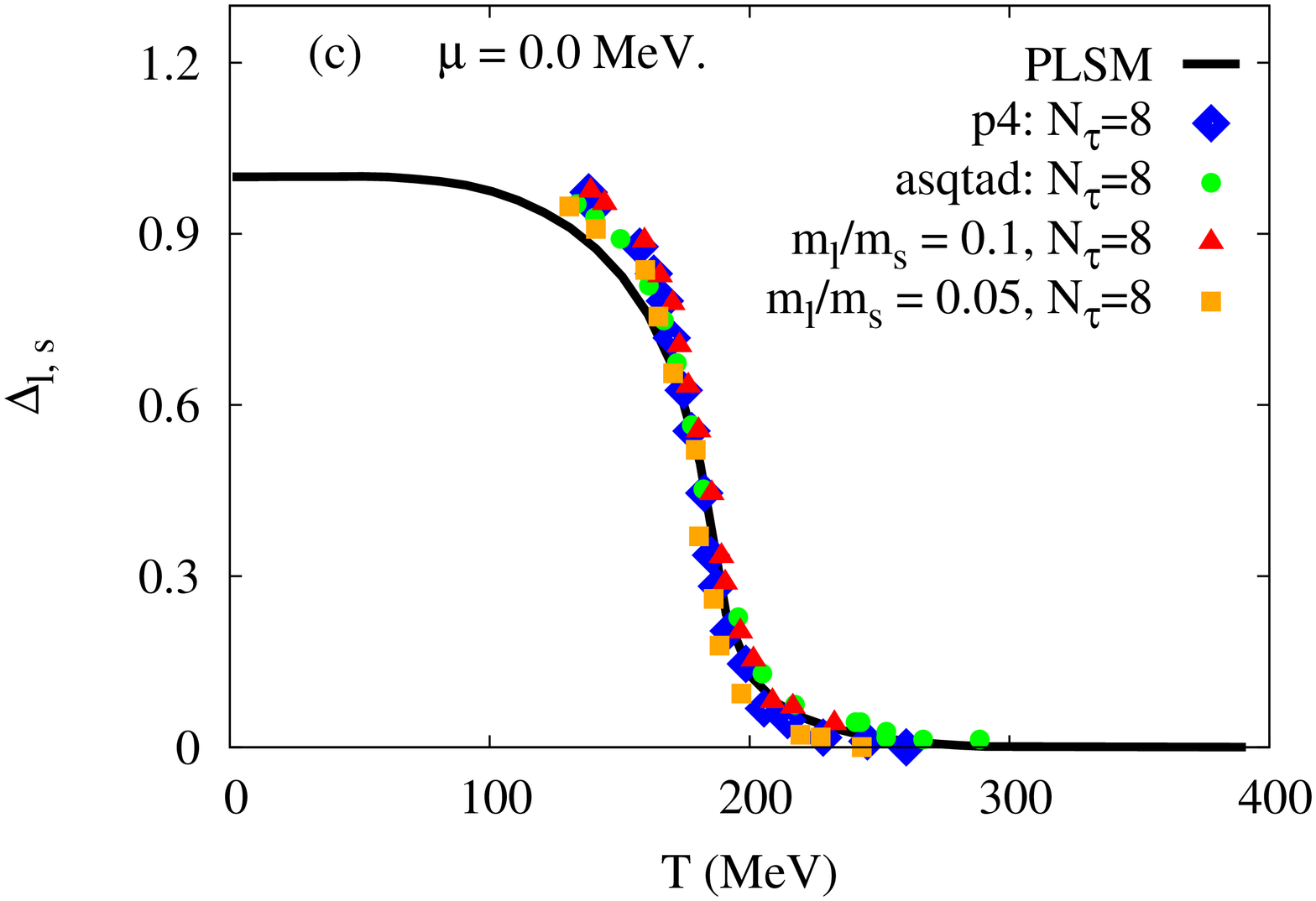}
\caption{\footnotesize Left-hand panel (a): Normalized chiral-condensates $\sigma _l/\sigma _{l_0}$ and  $\sigma_s/\sigma _{s_0}$  (solid and dotted curves) and deconfinement order-parameters $\phi$ and $\phi ^*$  (double-dotted and double-dotted dashed curves, respectively) calculated from logarithmic and polynomial-logarithmic Polyakov-loop potentials \cite{Roessner:2007, Sasaki:2013ssdw} are given as functions of temperatures at vanishing baryon chemical potential. Middle-panel (b): the same as left-hand panel but for the so-called chiral/deconfinement susceptibility, $\partial \sigma_f/ \partial T$ and $\partial \phi/ \partial T$. Right-hand panel (c) shows the subtracted condensates given as functions of temperature and compares the PLSM results with recent lattice QCD calculations \cite{lattice2009a, lattice2009b}.  \label{fig:cndst1}}
\end{figure}

In order to estimate the temperature dependence of the PLSM order parameters $\sigma_l$, $\sigma_s$, $\phi$, and $\phi^*$, the {\it real} part of thermodynamic potential, Eq. (\ref{cond1}), should be minimized, globally.  This is the procedure which is utilized in the present work. Alternatively, as the gap equations remain well-defined at finite $\mu$ we have solved the gap equations, directly; finding saddle point. Almost no difference was found. In doing this, we use $\sigma_{l_{0}}=92.4~$MeV and $\sigma_{s_{0}}=94.5~$MeV, which have been determined, experimentally \cite{Schaefer:2008hk,Mao:2010}. For details about the calculations of  $\sigma_l$, $\sigma_s$, $\phi$, and $\phi^*$ from PLSM, the readers are kindly advised to consult Refs. \cite{Tawfik:LSM,Tawfik:2014gga,Tawfik:Magnetic,Tawfik:quasi}. The results agree well with previous calculations, for instance, the temperature dependence of chiral and deconfinement order-parameters as shall be shown in Fig. \ref{fig:cndst1} with a special emphasize to the rapid decrease in the chiral condensates at the critical temperature.

In left-hand panel of Fig. \ref{fig:cndst1} (a), the normalized chiral condensates, $\sigma _l/\sigma _{l_0}$ and  $\sigma_s/\sigma_{s_0}$ which are correspondent to light and strange quarks, respectively, are given as functions of temperature.  At vanishing baryon chemical potential ($\mu=0$), the (thermal) expectation values of Polyakov-loop variable become identical, i.e., $\langle\phi\rangle=\langle\phi^{*}\rangle$ are determined. There is a noticeable difference between logarithmic \cite{Roessner:2007} and polynomial-logarithmic Polyakov-loop potentials \cite{Sasaki:2013ssdw}. It is apparent the the corresponding critical temperature considerably differs from logarithmic \cite{Roessner:2007} and polynomial-logarithmic expressions. These results give estimation for the critical temperatures.

Alternatively, the critical temperature can be relatively precisely estimated from the temperature-dependence of the susceptibility. What does {\it susceptibility} mean, shall be elaborated. Middle panel of  Fig. \ref{fig:cndst1} (b) shows the thermal evolution of the so-called {\it chiral susceptibilities},  $ \partial \sigma_f/ \partial T$ and $\phi_f/ \partial T$, where $\phi_f$ represents $\phi$ or $\phi^{*}$. This quantity gives the variation of a chiral quantity, such as $\sigma_f$ or $\phi_f$ with respect to the temperature, for instance. In literature, one used to refer to this quantity as {\it chiral/deconfinement susceptibility}. Peaks mark the rapid changes in the corresponding quantities as calculated from the Polyakov-loop potentials (logarithmic and polynomial-logarithmic). This would be taken as an estimation for the critical temperature, as well. It is obvious that the earlier refers to smaller critical temperature. As for $\sigma _l/\sigma _{l_0}$ and  $\sigma_s/\sigma_{s_0}$, the chiral critical temperature of light quarks is smaller than that of strange quarks, Tab. \ref{chiralT}.

For PLSM with $2+1$ quark flavors \cite{Schaffner:2013 chiral}, the subtracted condensates are calculated. Concretely, in our calculations $m_l$ and $m_s$ are substituted  by the symmetry breaking quantities $h_l$ and $h_s$, receptively,
\bea
\Delta_{l,s} = \frac{\left. \sigma _l - \left(\frac{h_l}{h_s}\right) \sigma_s \right|_T }{\left. \sigma_l - \left(\frac{h_l}{h_s}\right) \sigma_s \right|_{T=0}}.
\label{subtracted2}
\eea

The explicit symmetry breaking parameters $h_{i=l,s}$ can be - in turn - estimated from Dashen-Gell-Mann-Oakes-Renner (DGMOR) relations \cite{DGMOR1, DGMOR2}. They are directly related to nonstrange and strange flavors, on one hand. On the other hand, they are related to the pion and kaon masses, $m_\pi,\; m_K$, respectively \cite{Schaefer:2008hk}, 
\bea
h_l = f_\pi\, m_\pi^2,   \qquad  \qquad h_s = \sqrt{2} f_K m_K^2- \frac{f_\pi\,m_\pi^2}{\sqrt{2}},
\eea 
where $\bar{\sigma}_l=f_\pi$ and $\bar{\sigma}_s = (2f_K-f_\pi)/\sqrt{2}$. $h_{i=l,s}$ are fixed at different sigma-masses $m_\sigma$ \cite{Schaefer:2008hk}. $f_\pi$ and $f_K$ are vacuum decay constants of pion and Kaon, respectively. The parameters are listed out in Tab. \ref{tab:1a}.

The right-hand panel of Fig. \ref{fig:cndst1} (c) depicts the subtracted chiral condensates as functions of temperature at vanishing chemical potential. The PLSM calculations are compared with various $2+1$ lattice QCD simulations, in which asqtad \cite{Orginos:2009A,Orginos:2009B,Orginos:2009C} and p4 \cite{Heller:1999, Peikert:1998} improved staggered fermion actions with almost physical strange and light quark masses and temporal extent $N_\tau=8$ are implemented. The agreement between PLSM and lattice calculations is excellent. Accordingly, essential PLSM parameters can be determined. Thus, we conclude that the given parameters, especially, the three quark flavors are degenerate, model well the lattice QCD calculations.

It is obvious that $\Delta_{l,s}$ remains finite at low $T$. We observe that near $T_c$, $\Delta_{l,s}$ decreases very rapidly in a narrow range of temperatures, i.e., the light quark and gluon degrees-of-freedom liberate. In addition to these effects, the deconfinement phase-transition and/or the restoration of the broken chiral-symmetry can be characterized. 

Different studies have shown how ($2+1$) PQM model, where a quark-improved Polyakov-loop potential is included, leads to a smoother phase-transition between the hadronic phase (at low-temperature) and the quark-gluon plasma phase (at high-temperature) \cite{Schaffner:2013 chiral, Schaefer:2009ab}. Nevertheless, the modification of the gauge potential by the gluon potential in addition to the inclusion of quark-gluon interactions lead to a smoother and steeper decrease in the chiral phase-transition with increasing temperature. This seems to improve the agreement with the recent lattice QCD calculations. It is noteworthy highlighting that the adjustment of the pure-gauge potential to Polyakov-loop potential improves the calculations towards well-reproduction of the lattice QCD results \cite{lattice2009a, lattice2009b}. These modifications results in a smooth and steeper decrease in the chiral and deconfinement phase-transition(s) or crossover(s).

In order to estimate the chiral critical-temperature, two procedures can be implemented:
\begin{enumerate}
\item the first one is based on the intersection of the order parameters with the corresponding chiral condensates, right-hand panel of Fig. \ref{fig:cndst1} (a) and
\item the second one determines the temperature at which a peak of the chiral susceptibility ($\partial \sigma_f/ \partial T$ and $\partial \phi/ \partial T$) sets on, middle panel of Fig. \ref{fig:cndst1} (b).
\end{enumerate}
In the present work, we utilize the first method. 
It is believed that, the peak of the corresponding chiral susceptibility represents a better guide for the effective critical temperature. The values deduced from light- and strange-quark chiral-condensates, in which logarithmic, Eq. (\ref{LogULoop}), and polynomial-logarithmic Polyakov potentials, Eq. (\ref{LogPloy}), are utilized are listed out in Tab. \ref{chiralT}. We observe that the critical temperatures corresponding to strange quarks are greater than that to light quarks.

\begin{table}
\begin{center}
 \begin{tabular}{p{3cm} p{2.cm}  p{2.cm} p{2.cm} p{2.cm}} 
 \hline \hline
[MeV] & $T_{\chi}^l~$& $T_{\chi}^s~$ & $T_{\chi}^{\phi_{Log}}~$ & $T_{\chi}^{\phi_{PolyLog}}~$ \\  [0.5ex]  
 & $176.0$& $286.0~$ & $96.0$ & $186.0$ \\  
 \hline \hline
\end{tabular}
\end{center}
\caption{The chiral restoration temperatures determined at peaks of  $\partial \sigma_f/ \partial T$ and $\partial \phi/ \partial T$, middle panel of Fig. \ref{fig:cndst1}. \label{chiralT} }
\end{table}

Furthermore, the PLSM can be extended to determine the physical masses of the degenerated light and strange quarks under the assumption that the quark chemical potentials are equivalent, $\mu_u=\mu_d=\mu_s$. It is worthwhile to devote further efforts in determining the correlations and the fluctuations between the chiral and deconfinement phase-transition(s). According to the direct dependence of quark masses and their condensates,  $m_l = g \sigma_l/2$ and $m_s = g \sigma_s/\sqrt{2}$ and when taken into account the flavor-blind Yukawa coupling $g=6.5$, one can straightforwardly deduce that the mass of the light constituent quark $m_l \sim 300~$MeV and that of the strange constituent quark $m_s \sim 433~$MeV. This is a set of some PLSM parameters utilized in performing the present calculations.

\subsubsection{Relaxation time}
\label{sec:relxtime}

The precise estimation of the temperature dependence of the relaxation time ($\tau_f$) of QCD matter plays a very essential role in determining its conductivity and viscous properties in thermal medium. For example, this quantity is very crucial for the numerical estimation of the transport properties from Green-Kubo correlations \cite{Puglisi:2014,Reif:1965,Cassing:2014sai}. In framework of PLSM, the quark flavors represent the effective degrees-of-freedom, especially at high temperatures. Thus, the relaxation time of such a quark system, i.e., deconfined QCD matter, is strongly relevant to the present calculations. At lower temperatures, the hadronic degrees-of-freedom; the pion and sigma mesons, become dominant.  

\begin{figure}[htb]
\centering{
\includegraphics[width=7.5cm,angle=0]{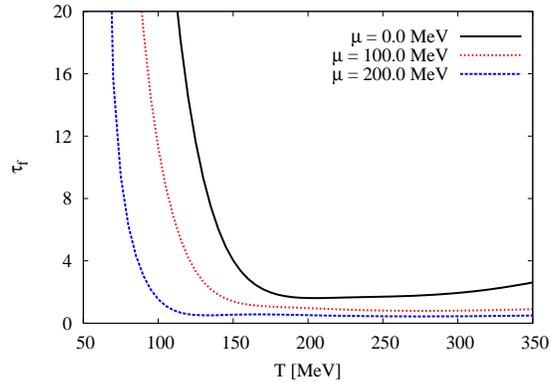}
\caption{\footnotesize The relaxation time (decay constant) of degenerate quarks, anti-quarks and gluons is given as functions of temperature at vanishing chemical potential $\mu=0$ (solid curve), $100$ (dotted curve) and $200~$MeV (dashed curve).
\label{decays}}}
\end{figure}
For a microscopic consideration, the relaxation time can be determined from the thermal average of total elastic scattering and depends on the relative cross-section $\sigma_{tr} (T)$,
\bea
\tau = \left[{n_f\, \langle \upsilon_{rel} (T) \, \sigma_{tr} (T) \rangle}\right]^{-1}, \label{RlaxTime}
\eea
where $\sigma_{tr}$ is the cross section, $\langle\upsilon_{rel}\rangle$ is the mean relative velocity of two colliding particles, and $n_f$ are their number densities. The relaxation time ($\tau$) has been evaluated from DQPM and NJL models \cite{Bratkovskaya}. In DQPM, $\tau$ is found indirectly-related to the finite decay-width of the quarks ($\Gamma$) and to that of the gluons; $\tau =\Gamma^{-1} $ \cite{Bratkovskaya}.  The temperature dependence of $\tau$ was introduced in Ref. \cite{Bratkovskaya}, in which DQPM was utilized and assumed that static quantities, such as, thermodynamics, are related to dynamic (flow) quantities, such as conductivity and viscosity.

For the sake of simplicity, the temperature dependence of $\sigma_{tr}$ can be determined according to the assumption of free (ideal) massless gas, where the confinement phase of mesons are conjectured to be liberated into deconfinement phase of free quarks and gluons, at very high temperatures, $T \gg T_c$. In ultra-relativistic kinetic theory,  the temperature dependence of the cross section ($\sigma_{tr}$) at temperatures around the critical value have a good agreement with the $T^{-2}$-dependence \cite{Greco:2009}.  Furthermore, from the Bjorken picture \cite{Armesto:2008, Molnar:20089}, we can assume that the temperature $T \sim \tau^{-1/3}$ and the cross section $\sigma_{tr} \approx \tau^{2/3}$. Then, we obtain that $\sigma_{tr} \sim T^{-2}$. Should these assumptions be correct, then the relaxation time can approximately be determined from the PLSM, where the temperature evolution of quark number-density becomes very obvious.

Fig. \ref{decays} shows the thermal dependence of the relaxation time ($\tau$) calculated from PLSM at different values of baryon chemical potential $\mu=0$ (solid curve), $100$ (dotted curve) and $200~$MeV (dashed curve).  This is the only figure in which we take into consideration the influence of the baryon chemical potential. It is obvious that the temperature dependence of $\tau$ rapidity decreases at low temperatures (hadron phase) and  remain almost unchanged at higher temperature (parton phase). On the other hand, increasing baryon chemical potential  considerably increases the decline in $\tau$. At higher temperatures, $\tau$ remains almost independent on the temperature.

As introduced in Ref. \cite{Zhuang95}, the transport theory with the Boltzmann master equation is used in calculating the momentum loss and the relaxation time. To this end, the screening of long-range quark-quark interactions is assumed for the process of inter-penetrating quark plasmas. For a spatially uniform quark plasma of favor $f$ flowing with respect to $N_f-1$ plasmas with another type of flavor, the relative flow velocity is assumed to relax in a long time due to the collision expansion. The resulting relaxation time can be utilized in calculating the transport coefficients from SU($3$) models \cite{Marty13}. 

It  was concluded that the relaxation rates for the momentum relaxation, the electrical conduction, and the viscous properties have a universal scale 
\bea
\tau &\simeq & \frac{m_q^{2/3}}{\left(\alpha_s\, T\right)^{5/3}},  \label{eq:tauT1}
\eea
where $\alpha_s$ is the running strong coupling and $m_q$ stands for quark mass, but due to the singular character of the interaction for small energy and small momentum transfer, the thermal conductivity behaves, differently  \cite{Zhuang95}. The relaxation rate for thermal conduction scales as 
\bea
\tau &\simeq & \frac{1}{\alpha_s\, T}. \label{eq:tauT2}
\eea

Alternative to Eqs. (\ref{eq:tauT1}) and (\ref{eq:tauT2}), it was assumed that the relaxation time in both partonic and hadronic phases  scales with the temperature $T$ as follow \cite{Zhuang95, Bratkovskaya}.
\bea
T<T_c: 
\left \{
\begin{array}{cc}
		    n\propto e^{-m/T}  	&  \\ & \\
		   \sigma \simeq \mbox{const.}	 &
\end{array}
\right. \Rightarrow \tau \propto e^{m/T}, \label{tausT}
\eea
and 
\bea
T>T_c: 
\left\{
\begin{array}{cc}
		    n\propto T^3 	&  \\ & \\
		   \sigma \propto T^{-2}&
\end{array}
\right. \Rightarrow \tau \propto T^{-1}. \label{taugT}
\eea
At low temperatures, the fast increase of the mass gives an exponential decrease in $\tau$ with increasing $T$ \cite{Bratkovskaya}. In this limit, $\sigma$ remains constant. At high temperatures, $m \ll T$, the cross section becomes proportional to $T^{-2}$ and $\tau$ linearly decreases with increasing $T$.

To answer the question why models used to study static properties of the QCD phases (they might not include even the dynamics of the phase transition) can be used in studying the transport properties, and to what extent possible defects in these models can be excused?, 
we propose that the "relaxation time",  plays the role of a translator from ''static'' to ''transport'' quantities and thus enables PLSM to reproduce first-principle lattice calculations on both ''static'' and ''transport'' properties. In other words, the excellent reproduction of recent lattice QCD results should be understood due to the excellent estimation of the various PLSM parameters and the good modelling of the relaxation time. To summarize, the relaxation time was modelled in Eqs. (\ref{tausT})-(\ref{taugT}). This modelling is not closely depending on PLSM.

First, we  have evaluated the PLSM order-parameters as functions of temperatures by minimizing the real potential, at a saddle point. Then, we have calculated the dependence of the relaxation time on temperature, Eq. (\ref{RlaxTime}), for which, the quark number should be calculated by defining an effective distribution function and energy-momentum dispersion relation where quark masses are coupled to the sigma fields $\sigma_l$ and $\sigma_s$ for light and strange quarks, respectively. As can be taken from Eqs. (\ref{electric_cond}), (\ref{heatcond}), (\ref{zetaTMU}), and (\ref{etaTMU}), the relaxation time is an essential quantity in calculating all these transport properties.

\subsection{Conductivity properties \label{conductivitya}}

\subsubsection{Electrical conductivity} 

The electrical conductivity ($\sigma _{el}$) is a key transport coefficient, which recently gains an increasing interest among the particle physicists. This property is related to the flow of the charge carriers, especially in presence of an electrical field. To measure the electrical conductivity, there are two methods suggested so far.
\begin{itemize}
\item The first one applies an empirical methodology. An external electrical field is applied to the system of interest. If a small uniform electric field ($\mathcal{E}$) is applied to the $z$-direction, a nonequilibrium situation results in orienting the electrical current density in that direction ($j_z$). By definition $j_z = \sigma _{el}  \mathcal{E}$, where the proportionally constant ($\sigma _{el}$) is called the electrical conductivity \cite{Puglisi:2014}, this relation is called Ohm's law. 
\item The second one refers the self-interaction between quarks and gluons, i.e., no external electrical field is needed. The relativistic motion of the electrically charged quarks play this role. The electrical conductivity is related to the electric flow of charged quarks.
\end{itemize}
Consider light and strange quark flavors having mass ($m$) and elementary charge ($e$) interacting with each others, their collision time ($\tau$) is determined from their scattering. When an electrical field (${\vec{\mathcal{E}}}= \mathcal{E} \hat{e}_z$) is applied in the $z$-direction, it gives rise to a mean $z$-component of the velocity ($\bar{v}_z$) of the charged quark flavors. The electric current density is equal to the mean number ($n$) crossing a unit area perpendicular to the $z$-direction per unit time multiplied by $e$ charges, i.e., $n e \bar{v}_z$, 
\bea
j_z = n\,e \,\bar{v}_z.
\eea
In a system of charged particles, one obtains that, $\bar{v}_z=  e \mathcal{E} \tau /m$ \cite{Reif:1965}. 

Furthermore, it is noteworthy highlighting that the electrical conductivity is a well known property of classical gas. At finite temperature ($T$) and baryon chemical potential ($\mu$),  Drude-Lorentz conductivity  can be expressed as \cite{Reif:1965, Heiselberg1993}
\bea
\sigma_{el} = \sum_{f} e_f^2\, \frac{n_f(T, \mu)\,\tau _f (T,\, \mu)}{m_f (T, \mu)},
\label{electric_cond}
\eea 
where $f$ runs over quarks ($u$, $d$ and $s$) antiquarks ($\bar{u}$, $\bar{d}$ and $\bar{s}$) and gluons $g$ flavors. The electric charge of quarks are summed up as $\sum q_f$. In all these estimations, free space is assumed. Thus, the electromagnetic coupling or the fine-structure constant at zero energy is $\alpha=1/137$ \cite{ele:B_Amato:2013}. The corresponding lattice QCD calculations are normalized by  $5/9$ for $2$ quark-flavors  \cite{ele:B_Amato:2013}. Nevertheless, the electric charge of quarks is given as \cite{Bratkovskaya}, 
\bea
e_f^2 = \frac{4\pi}{137} q^2,
\eea 
where $q$ is the quark electric-charge fraction, $q=+2/3$ or $-1/3$. 

The functions $n_f(T, \mu)$, $\tau_f(T, \mu)$ and $m_f(T, \mu)$ stand for number density, collision time or relaxation time and  the corresponding mass, respectively, at finite $T$ and finite $\mu$. In all  present calculations, we assume vanishing baryon chemical potential. Equation (\ref{electric_cond}) indicates that the electrical conductivity is related to thermodynamic properties besides the relaxation time and the effective mass. As discussed in earlier section, having a good model for the temperature dependence of the relaxation time is very essential in enabling PLSM to reproduce lattice calculation on both static (thermodynamic quantities, for example) and transport properties.

The left-hand panel of Fig. \ref{conductivity} (a) shows the temperature dependence of the electrical conductivity ($\sigma_{el} /T$) at vanishing chemical potential ($\mu=0$). The normalized electrical conductivity calculated from PLSM is compared with difference lattice QCD calculations \cite{ele:A_Gupta:2004, ele:B_Amato:2013, ele:c_Francis:2011, ele:D_Francis:2013} and with other  QCD-like models such as NJL and DQPM \cite{Bratkovskaya}, where the circles denote lattice size $N_{s}=24^3$, while the square points are calculation on $N_{s}=32^3$ lattice, Fig. \ref{conductivity}. Both calculations assume $2+1$ quark flavors, but the lattice QCD  given as closed circles \cite{ele:D_Francis:2013}, crosses \cite{ele:c_Francis:2011} and open triangles  \cite{LQCD:ratioA,LQCD:ratioB,LQCD:ratioC} are zero-flavored. 

We observe that the PLSM results agree well with the lattice QCD calculations \cite{ele:B_Amato:2013}, especially at $T>T_c$. The difference between both lattice QCD simulations might be originated to the the different methods of simulations implemented on both of them. The PLSM electrical conductivity curve, which is calculated from Eq. (\ref{electric_cond}), refers to a combination between the quark number multiplicity and their masses. DQPM and NJL results \cite{Bratkovskaya} fairly agree with the lattice QCD  \cite{ele:B_Amato:2013}, especially at $T<T_c$.

From the temperature dependence of the normalized electrical conductivity calculated from PLSM and the comparison with other QCD-like models, such as NJL and DQPM \cite{Bratkovskaya}, we conclude that the PLSM results are most comparable with the lattice simulations \cite{ele:B_Amato:2013} and accordingly the PLSM parameters enable this QCD-like approach, in which Polyakov-loop potentials are integrated, to reproduce first-principle lattice calculations. Furthermore, we conclude that, the proposed temperature dependence of the relation time, section \ref{sec:relxtime}, and that of the chiral condensates and of the deconfinement order-parameters, section \ref{structure}, are - besides other PLSM parameters - essential in empowering SU($3$) PLSM to this reliable production of the lattice QCD simulations.

\subsubsection{Heat conductivity}

The heat conductivity, $\kappa (T, \mu)$, is related to the heat flow in relativistic fluid \cite{Israel_Stewart, Groot:1980, Rischke1, Greiner:2013} and gives an indicator about the rate of energy change taking place in the system of interest. A simple way to estimate the time evolution of the heat conductivity is the simulation of the likely irradiation occurring in the system  through energetic ions \cite{heat_conductivity}. From specific heat ($c_v$) and relaxation time ($\tau$), the heat conductivity reads \cite{Heiselberg1993},
\bea
\kappa (T, \mu) = \frac{1}{3} \nu_{rel}\, c_v (T, \mu) \sum_f \tau_f(T, \mu), \label{heatcond}
\eea 
where $\nu_{rel}$ is the relative velocities. All quantities in Eq. (\ref{heatcond}) are thermal averages. For simplicity, we  assume that  $\nu_{rel}\sim 1$.  For two quarks with masses $m_1$ and $m_2$, respectively, the center-of-mass collisions result in the relative velocity $\nu _{rel} = \sqrt{(p_1 p_2)^2 - (m_1 m_2)^2}/E_1 E_2$ \cite{Sasaki:2010}. In the relativistic limit, the quark masses are negligibly small relative to the momentum, where the quark masses decrease with increase $T$ according to the chiral condensate, Fig. \ref{fig:cndst1}. From Eq. (\ref{heatcond}), the heat conductivity is strongly related to the decay constant or the relaxation time of the quarks which - in tern - depends on the temperature and the baryon chemical potential. 

\begin{figure}[htb]
\centering{
\includegraphics[width=5.5cm, height=4.cm, angle=-0]{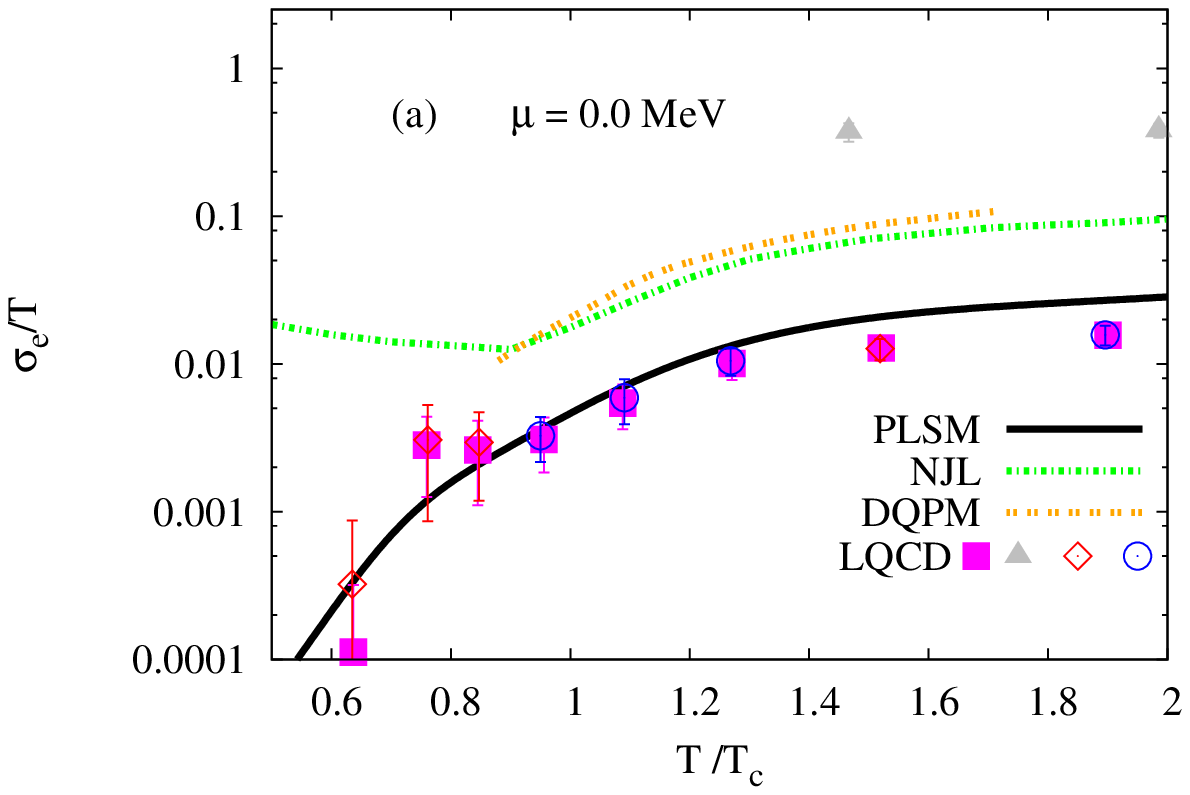}
\includegraphics[width=5.cm, height=4.cm,angle=-0]{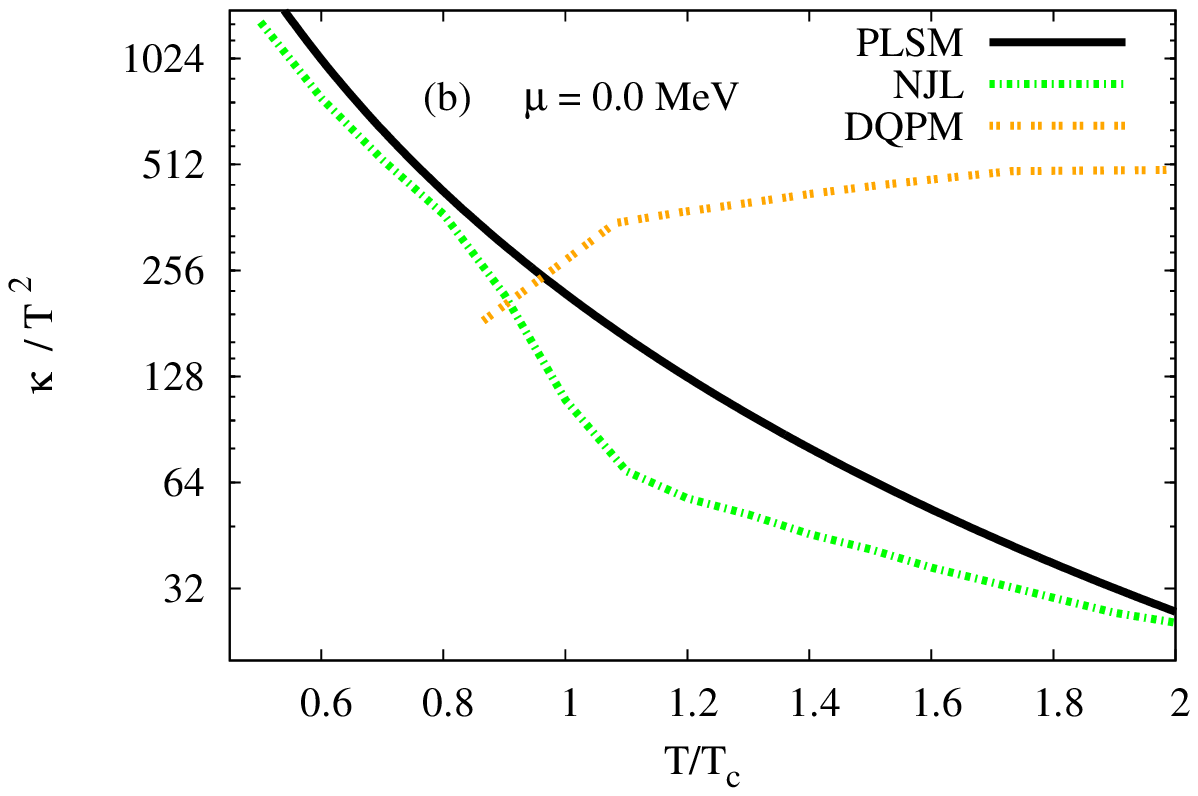}
\includegraphics[width=5.cm, height=4.cm,angle=-0]{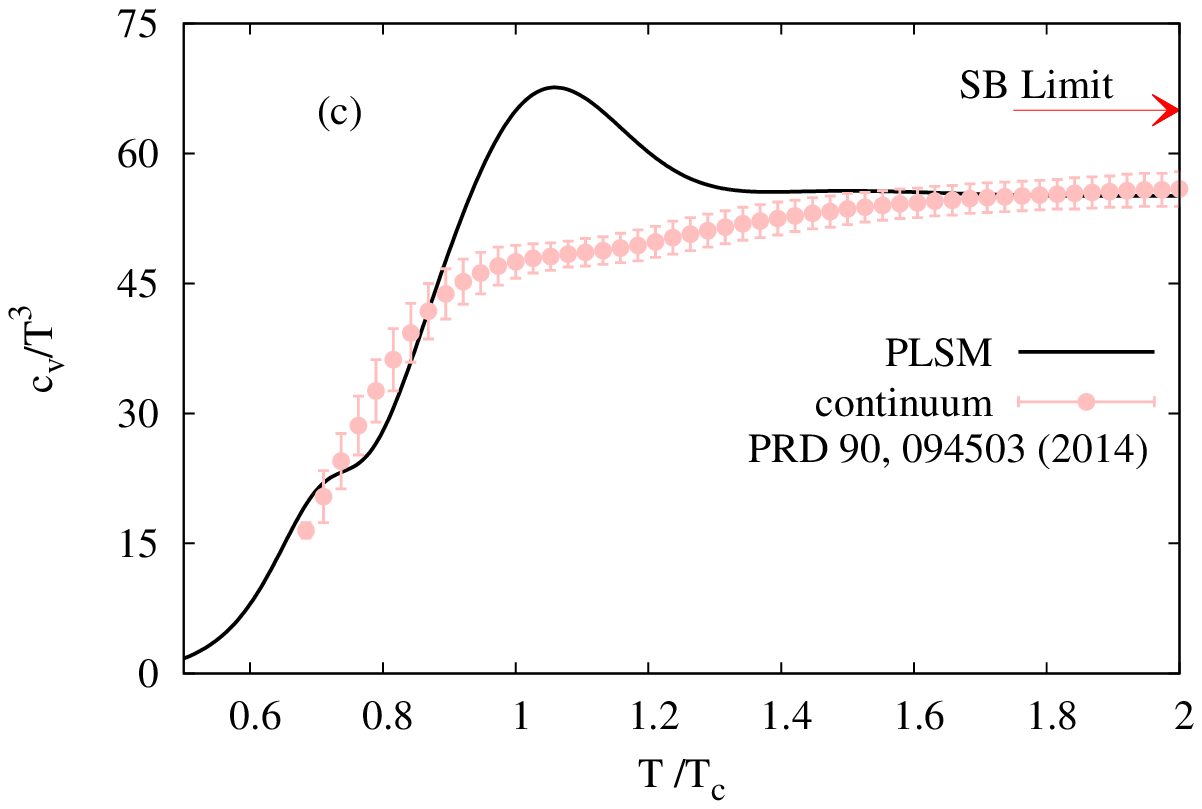}
\caption{\footnotesize Left-hand panel (a): the normalized electrical conductivity as a function of temperature at vanishing chemical potential is calculated from PLSM (solid curve) and compared with NJL \cite{Bratkovskaya} (dotted dash) and DQPM  \cite{Bratkovskaya} (double dotted) and lattice QCD simulations \cite{ele:B_Amato:2013} (circle points), (square points), \cite{ele:D_Francis:2013} (closed circle points), \cite{ele:c_Francis:2011} (cross point) and \cite{LQCD:ratioA,LQCD:ratioB,LQCD:ratioC} (open triangle). Middle-panel (b): the heat conductivity normalized to $T^2$ is calculated as a function of temperature at vanishing chemical potential  calculated from PLSM (solid curve) and compared with NJL  \cite{Bratkovskaya} (dotted dash) and DQPM (double dotted) \cite{Bratkovskaya}. Right-hand panel (c): the dependence of the normalized specific heat ($c_v/T^3$) to temperature is compared with recent lattice QCD simulations \cite{HotQCDLattice:2014}.    
\label{conductivity}}}
\end{figure}

The middle-hand panel of Fig. \ref{conductivity} (b) shows the heat conductivity normalized to $T^2$ as a function of temperature at vanishing baryon chemical potential. The PLSM results are compared with NJL and DQPM calculations \cite{Bratkovskaya}. It is worthwhile to highlight that the different models have different critical temperatures, $T_c \sim 240~$MeV from PLSM,  $T_c \sim 200~$MeV from NJL and  $T_c \sim 158~$MeV from DQPM. The temperature dependence of NJL heat-conductivity normalized to $T^2$, Eq. (\ref{heatcond}), decreases faster than the one from PLSM \cite{Bratkovskaya}. From DQPM, the temperature dependence is the opposite. Here, increasing temperature increases the heat conductivity. There are no lattice QCD calculations to compare with them.  Equivalently, we compare our calculations on specific heat ($c_v$) with recent lattice QCD calculation \cite{HotQCDLattice:2014}. This is given in the right-hand panel (c). We find an excellent agreement, especially at low and high temperatures. At temperatures around $T_c$, a peak is positioned. It is apparently located at $T_c$. This can be interpreted from the definition of the specific heat;  $c_v =\partial \epsilon/\partial T$, where the rapid change in the energy density ($\epsilon$) around $T_c$ region gives a plausible interpretation of the observed peak in $c_v$. Furthermore, this peak most-probably depends on the baryon chemical potential ($\mu$), not shown here. In a previous work, we have shown that the peak seems to decrease with increasing the baryon chemical potential \cite{TN:thermo2015}.

\subsection{Viscosity properties}
\label{Viscosity}

In the present work, the transport properties of QCD matter in thermal medium is limited to bulk and shear viscosities. These are strongly related to the hydrodynamical flow of relativistic fluid, QGP in our case, and the transverse motion of its constituents (partons) during the expansion of the strongly interacting system \cite{Stephanov:2006, Stephanov2008}. In other words, the estimation of viscosity is very crucial to characterizing the evaluation of essential physical observables such as the elliptic flow ($v_2$)  \cite{Stephanov:2006, Stephanov2008} and the correlation functions \cite{Stephanov:2006, Stephanov2008}. 

Corresponding to the dissipative fluxes, the Green-Kubo (GK) correlations, which are based on the linear response theory (LRT)  \cite{Fraile:2009}, directly relate the transport coefficients to out- and in-equilibrium correlations. It is noteworthy to emphasize that the calculation of the transport coefficients, for instance, are based on linear response to perturbation in disturbance but not perturbation in coupling. Linear response is what allows the extraction of non-equilibrium information from correlators, which are calculated in equilibrium. The effective theory is believed to model the medium in a very different way than that in lattice QCD. Although, the lattice computation of these correlators is not the same as in an effective model, the reproduction of the first-principle calculations becomes strong evidence that certain class of correlators are correctly estimated and the parameters chosen for QCD-like approach are correctly determined.

The dissipative fluxes are treated as perturbations to local thermal-equilibrium. In doing this, the transport coefficients associated with conserved quantities can be formulated as expected values at equilibrium \cite{Fraile:2009}. According to Green-Kubo (GK) correlations, the bulk and shear viscosities are given in Lehmann spectral representation of two-point correlation functions as the components of the energy-momentum tensor, such as \cite{Kubo:1957}
\bea
\left(
\begin{array}[c]{c}
\zeta  \\ \eta
\end{array}
\right) 
=
\lim_{\omega\rightarrow 0^+} \lim_{|{\bf p}|\rightarrow 0^+} \frac{1}{\omega}
\left(
\begin{array}[c]{c}
\frac{1}{2} A_{\zeta} (\omega, |{\bf p}|) \\ \frac{1}{20} A_{\eta} (\omega, |{\bf p}|) 
\end{array}
\right), \label{eq:matrix_field_A}%
\eea
where $A_{\zeta}$  and $ A_{\eta}$ are spectral functions \cite{Kubo:1957}
\bea
A_{\zeta} (\omega, |{\bf p}|)&=& \int d^4 x\; e^{ip\cdot x} \langle\left[\mathcal{P}(x), \mathcal{P}(0)\right] \rangle, \\ 
A_{\eta} (\omega, |{\bf p}|) &=& \int d^4 x\; e^{ip\cdot x} \langle\left[\pi^{ij}(x), \pi^{ij}(0)\right] \rangle,
\eea
with  
\bea
\mathcal{P}(x) &=& -\frac{1}{3} T^{i}_{i} (x) - c_s^2 T^{00} (x), \\
\pi^{ij}(x)  &=& T^{ij} (x) -\frac{1}{3} \delta^{ij} T^{k}_{k} (x),
\eea
and $\langle\left[ \cdots \right] \rangle$ donates an appropriate thermal average. 
The lowest-order contributions to the bulk and shear viscosity,  respectively,  read  \cite{Marty13,Kapusta2011},
\bea
\zeta (T, \mu) &=& 12  \sum_f  \int \, \frac{1}{T}\, \frac{d^3p}{(2\pi)^3} \, \frac{\tau_f}{E_f^2} \left[\frac{|\vec{p}|^2}{3} - c_s^2\, E_f^2 \right]^2 \, n_{f} (T, \mu) \Big[1 - n_{f} (T, \mu)  \Big], \label{zetaTMU} \\ 
\eta (T, \mu) &=& 12  \sum_f \int \, \frac{1}{15\, T}\, \frac{d^3p}{(2\pi)^3} \, \frac{|\vec{p}|^4 \tau_f}{E_f^2} \, \, n_{f} (T, \mu) \Big[1 - n_{f} (T, \mu)  \Big], \label{etaTMU}
\eea  
where $E_f$ is the single-state energy. The factors in rhs of both expressions count for $N_c=3$, spin, and particle-antiparticle degeneracies.
When introducing the Polyakov-loop potentials to LSM, Eq. (\ref{LogPloy}), Fermi-Dirac distributions get modifications  \cite{Marty13,Kapusta2011}, 
\begin{eqnarray}
n_f &=&
\frac{\left(\phi^*+2 \phi \,e^{-\frac{E_f-\mu _f}{T}}\right)\, e^{-\frac{E_f-\mu _f}{T}}+e^{-3 \frac{E_f-\mu _f}{T}}}{1+3\left(\phi^*+\phi \,e^{-\frac{E_f-\mu _f}{T}}\right)\, e^{-\frac{E_f-\mu _f}{T}}+e^{-3 \frac{E_f-\mu _f}{T}}},
\label{fqaurk} 
\end{eqnarray}
where the corresponding functions in Eq. (\ref{PloykovPLSM}) are identical. For antiquarks, $\phi$ is to be replaced by $\phi^*$ and the chemical potential $\mu$ by $-\mu$. It is obvious how to assure that, in the deconfined phase, $\phi$, $\phi^* \rightarrow 1$ and Eq. (\ref{fqaurk}) approaches the standard definition. When the quarks are confined in colorless bound states, $\phi$, $\phi^* \rightarrow 0$ and the Boltzmann exponent will be multiplied by a factor $3$.  
At zero Polyakov loops, the emergence of factor $3$ in the Boltzmann exponent assures a statistical confidence that only $3$-quark states and not $1$- or $2$-quark states are allowed in the statistical sum for the partition function.

The quenched lattice study of Euclidean energy-momentum tensor ($T_{\mu}^{\mu}$) correlation-function has been performed in Ref. \cite{Meyer:2007A,Meyer:2007B}. The author aimed to introduce an analytic continuation of the lattice simulation in order to determine the real-time spectral-function and bulk viscosity. The analyse of the spectral function [$\rho(\omega)$] in a weak coupling regime and close to the phase transition  are essential quantities. The bulk viscosity is determined as $\xi=(1/2) \lim_{\omega\rightarrow 0} \rho(\omega)/\omega$  \cite{rff66}. At weak coupling, i.e., very high energies, which is related to vanishing chemical-potential, the spectral function can be calculated, perturbatively. Near second-order phase-transition as found in realistic QCD, especially at the critical endpoint, scaling behaviors can be determined. In both regime, the authors of Ref. \cite{rff66} proposed to subtract perturbative contributions and by replacing the spectral function with its thermal part, the vacuum energy contribution can be removed.

\subsubsection{Bulk and shear viscosities}

\begin{figure}[htb]
\centering{
\includegraphics[width=6.5cm,angle=-0]{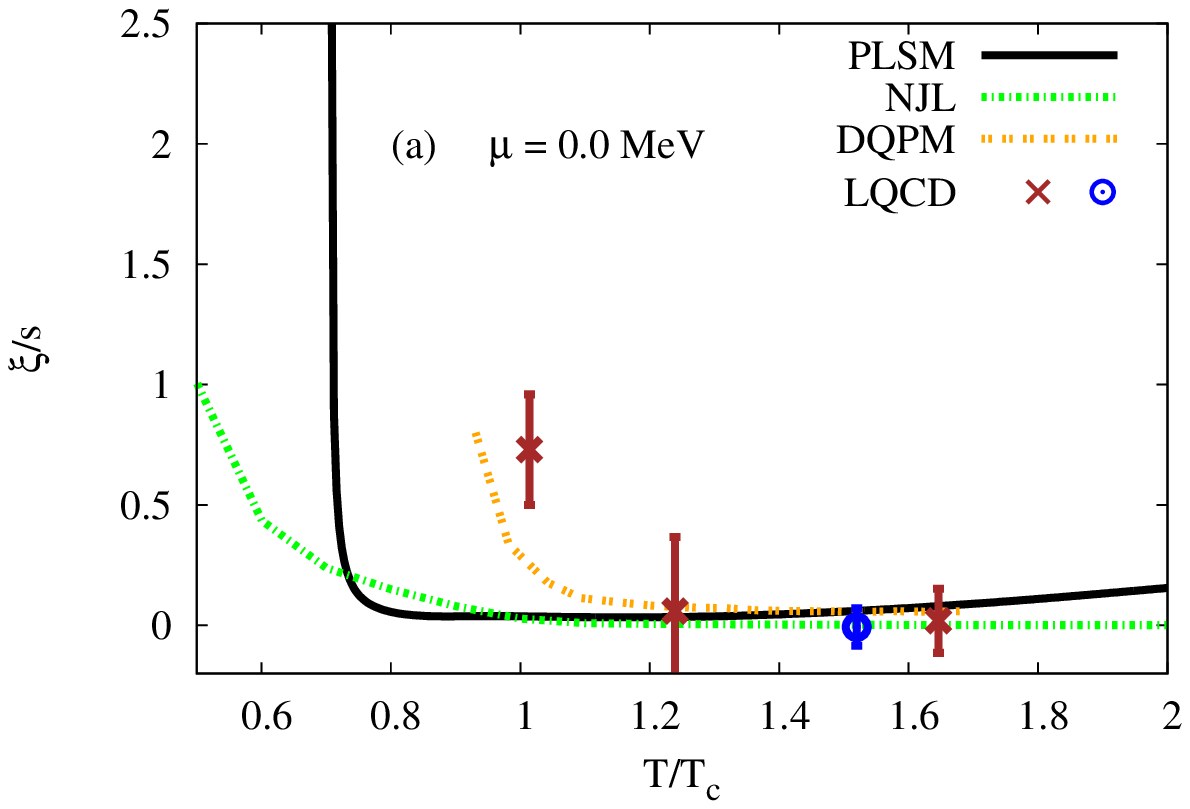}
\includegraphics[width=6.5cm,angle=-0]{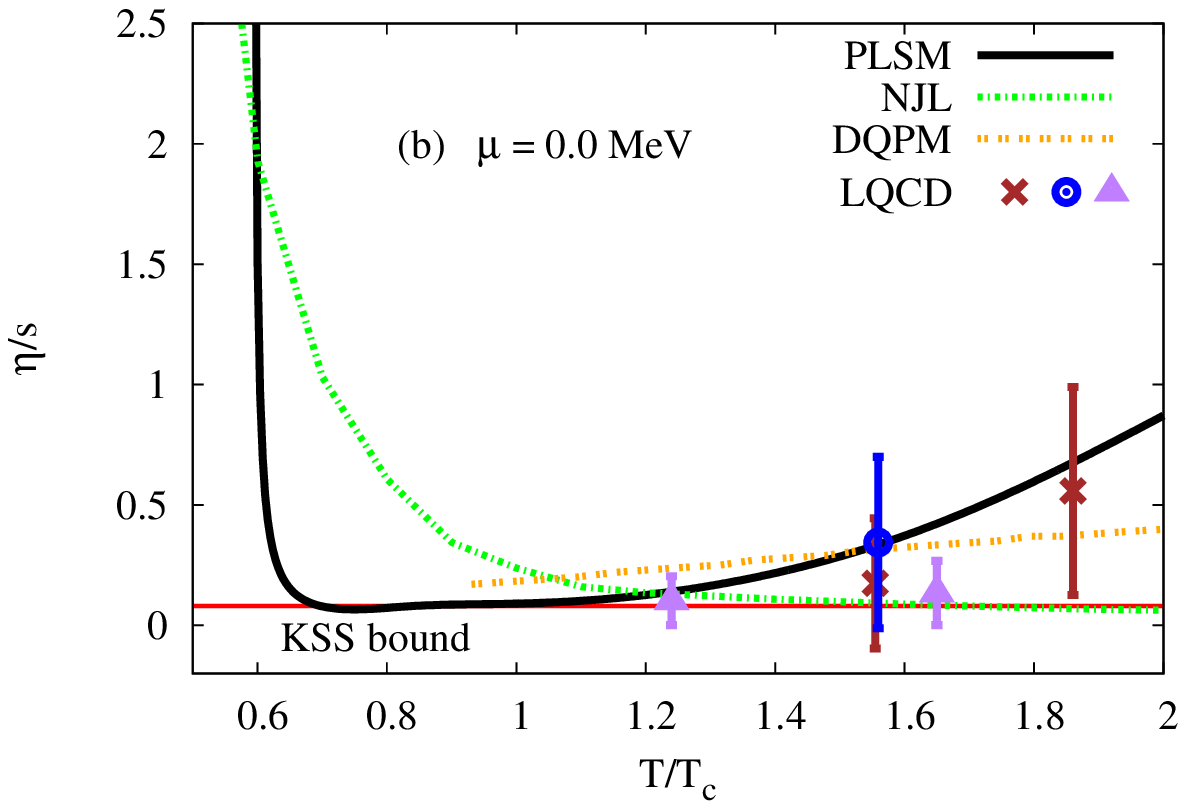} 
\caption{\footnotesize Left-hand panel (a): the ratio of bulk viscosity and thermal entropy ($\xi/s$) calculated from PLSM (solid curve) and compared with lattice QCD simulations \cite{Meyer:2007A,Meyer:2007B} (cross points) and \cite{Sakai:2007} (circle points) and (square points) is given as a function of temperature at vanishing chemical potential. Right-hand panel (b): shows the ratio of shear viscosity to thermal entropy ($\eta/s$) calculated from PLSM (solid curve) and compared with the lattice QCD  simulations  \cite{Meyer:2007A,Meyer:2007B} (cross points), \cite{Sakai:2007} (circle points) and (square points)  and   \cite{Sakai:2005} (triangle points). The Kovtun-Son-Starinets (KSS) bound is drawn. In both panels the results are compared with  NJL (dotted dash) and DQPM (double dotted) \cite{Bratkovskaya}. 
\label{viscosity}}}
\end{figure}

In left-hand panel of Fig. \ref{viscosity} (a), the temperature evolution of the ratio of bulk viscosity $\xi$ to the thermal entropy $s(T)$ at vanishing chemical potential is presented. The ratio of shear viscosity to the thermal entropy ($\eta/s$) as function of temperature at vanishing chemical potential is given in the right-hand panel. At temperatures close to the critical one, $\xi /s$ shows a good agreement with the lattice QCD calculations \cite{Karsch:2008,Meyer:2007A,Meyer:2007B,Sakai:2007,Sakai:2005}. The agreement with LSM \cite{Mao:2010, Paech:2006}, DQPM and NJL  approaches \cite{Bratkovskaya} are good, as well.  It seems that the entropy tends to vanish in order to decrease the temperature.  Around $T_c$, $\xi /s$ rapidly decreases. The sharp increase in bulk viscosity is positioned near the phase transition or seems to induce instability in the hydrodynamic flow of the formed QGP. This might be responsible for some RHIC observables \cite{Torrieri:2008A,Torrieri:2008B}. Thus, investigating $\xi/s$ would have a great impact on some experimental observables.  

The shear viscosity is a very suitable quantity to understood the phase transition between hadrons and quarks. A good agreement with lattice QCD calculations \cite{Karsch:2008,Meyer:2007A,Meyer:2007B,Sakai:2007,Sakai:2005} and other QCD-like effective models \cite{Bratkovskaya,Bratkovskaya:2013} is observed, right-panel of Fig. \ref{viscosity} (b).  In particular, the lower values of the ratio of shear viscosity to entropy refers to low QGP viscosity in the partonic phase \cite{KhvorostukhinA,KhvorostukhinB}. Such decrease is caused by the stronger interactions and the effective degrees-of-freedom. It is supported by the experimental description of the collective flow in heavy-ion collisions \cite{Cassing:2008, Konchakovski:2012A,Konchakovski:2012B}. It is noteworthy noticing that the numerical estimation of viscosity normalized to entropy from PLSM is higher than KSS bound \cite{Kovtun:2005}, which is $T$-independent, $\sim 1/4\pi$.

\subsection{Comparison between conductivity and viscous coefficients}

\begin{figure}[htb]
\centering{
\includegraphics[width=5.cm,angle=-90]{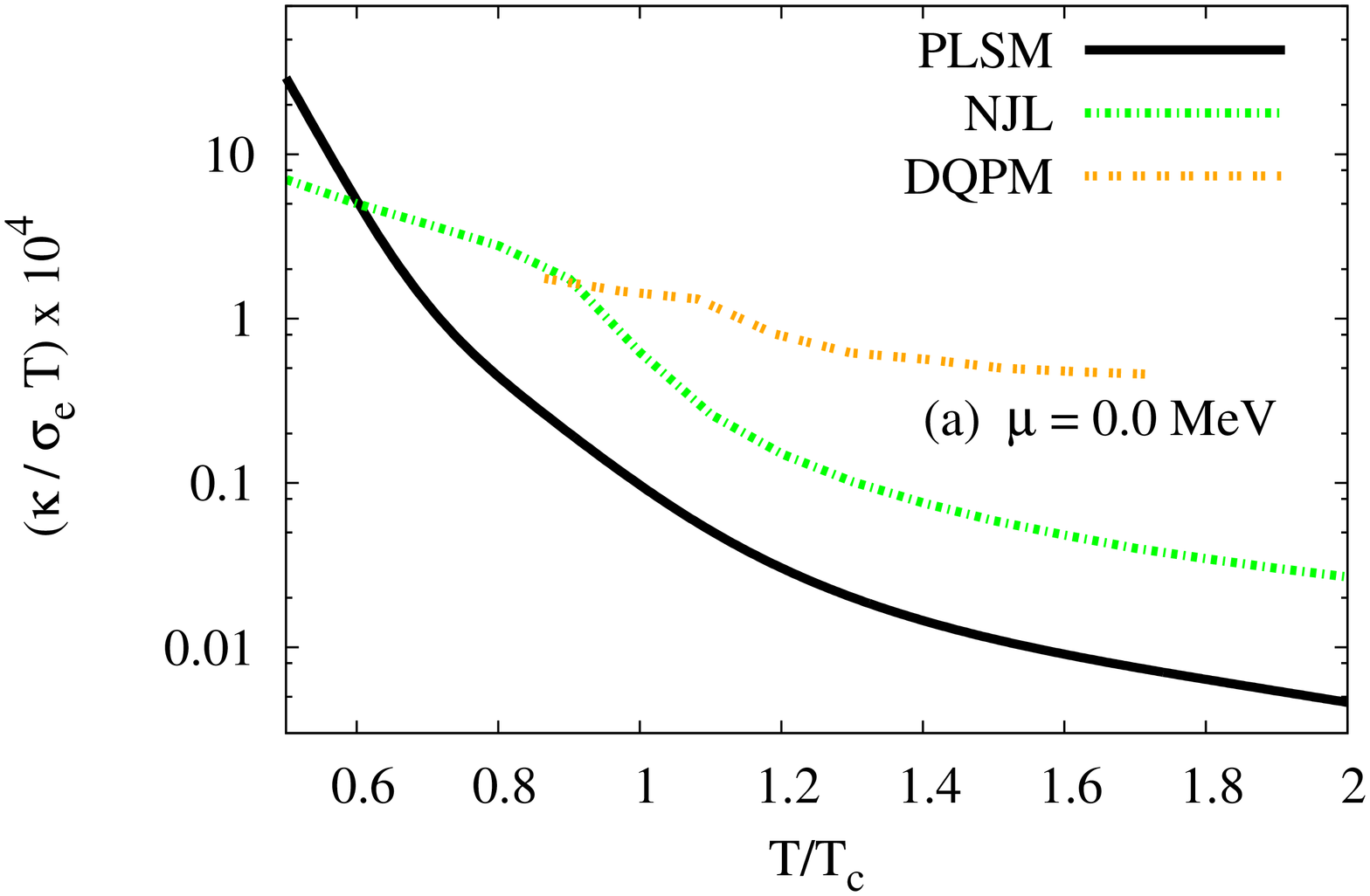}
\includegraphics[width=5.cm,angle=-90]{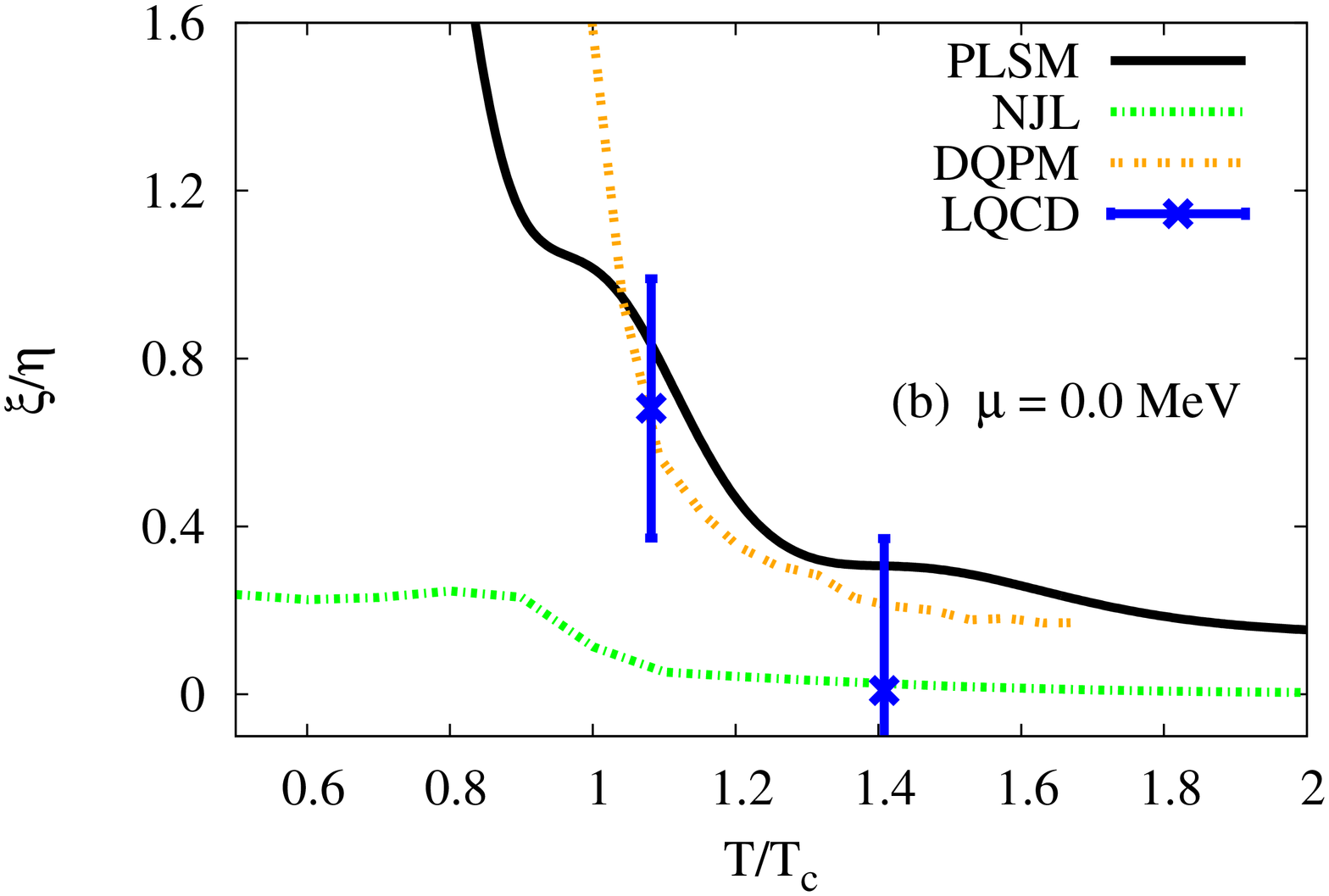}
\caption{\footnotesize Left-hand panel (a): numerical estimation for thermal (heat) to electrical conductivities ratio as a function of temperature at vanishing chemical potential is compared with QCD-like models, PLSM (solid curve), NJL (dotted dash) and DQPM (double dotted) \cite{Bratkovskaya}. Right-hand panel (b): the ratio of bulk to shear viscosities at vanishing chemical potential is compared with  lattice QCD calculations \cite{Meyer:2007A,Meyer:2007B}
\label{ratio_cond1_vis}}}
\end{figure}

The left-hand panel of Fig. \ref{ratio_cond1_vis} (a) presents the numerical estimation for $\kappa/T^2$-to-$\sigma_e /T$ ratio calculated from different effective SU($3$) approaches. It is obvious that the PLSM results rapidly decrease. They are faster than the ones calculated from NJL and DQPM \cite{Bratkovskaya}, especially at temperatures exceeding the critical one ($T_c$). There are no lattice QCD calculations to be compared with the present calculations. At $T>T_c $, the ratios from the different SU($3$) approaches are distinguishable by about one order of magnitude.

The right-hand panel of Fig. \ref{ratio_cond1_vis} (b) shows the ratio of bulk to shear viscosities ($\eta/\xi$) as a function of temperature. When the temperature approaches $T_c$, a sudden decrease takes place and the ratio tends to be temperature-independent. This behavior seems confirmed by recent lattice QCD calculations \cite{Meyer:2007A,Meyer:2007B}. 

\begin{figure}[htb]
\centering{
\includegraphics[width=5.5cm,angle=-90]{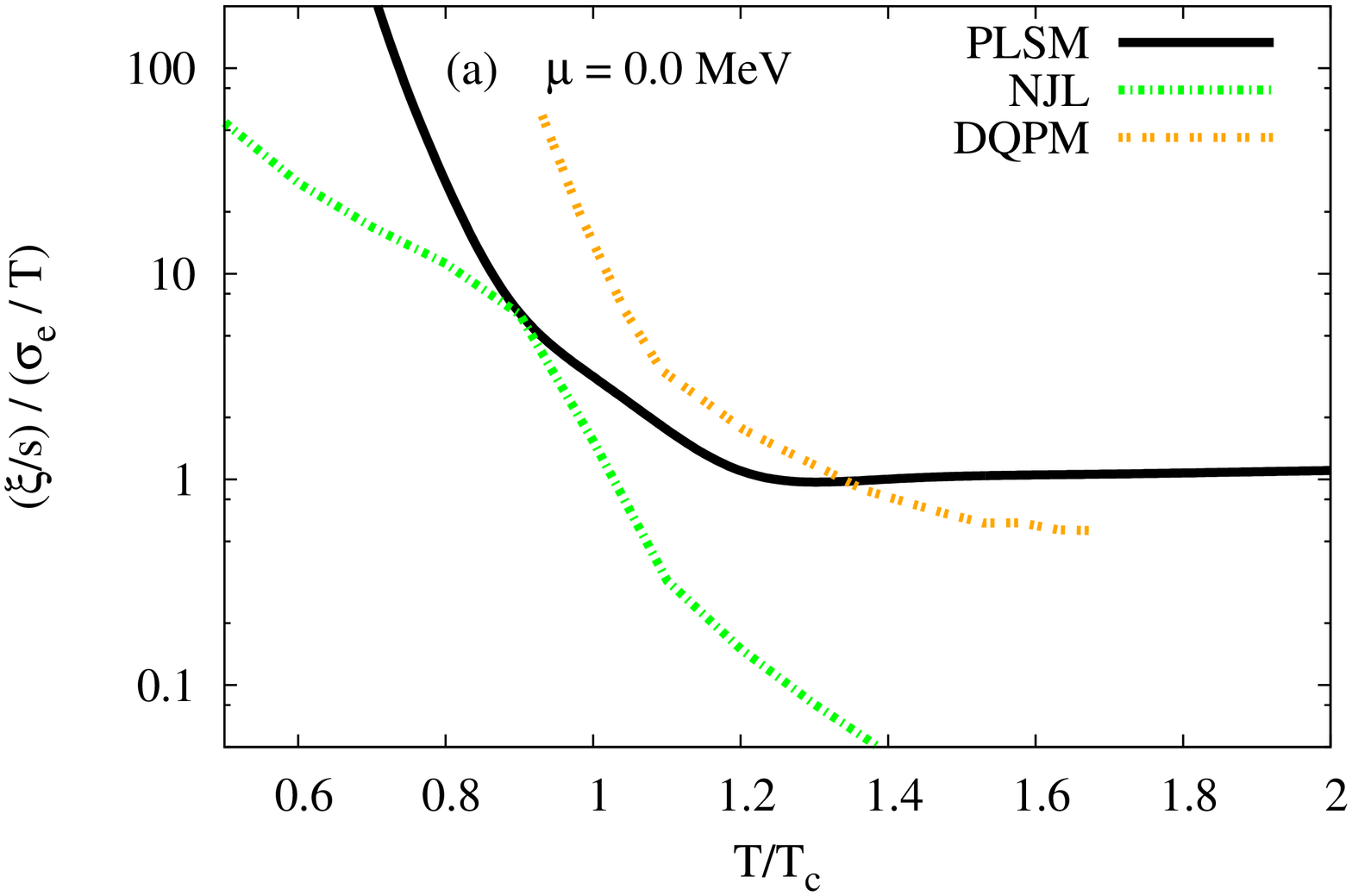}
\includegraphics[width=5.5cm,angle=-90]{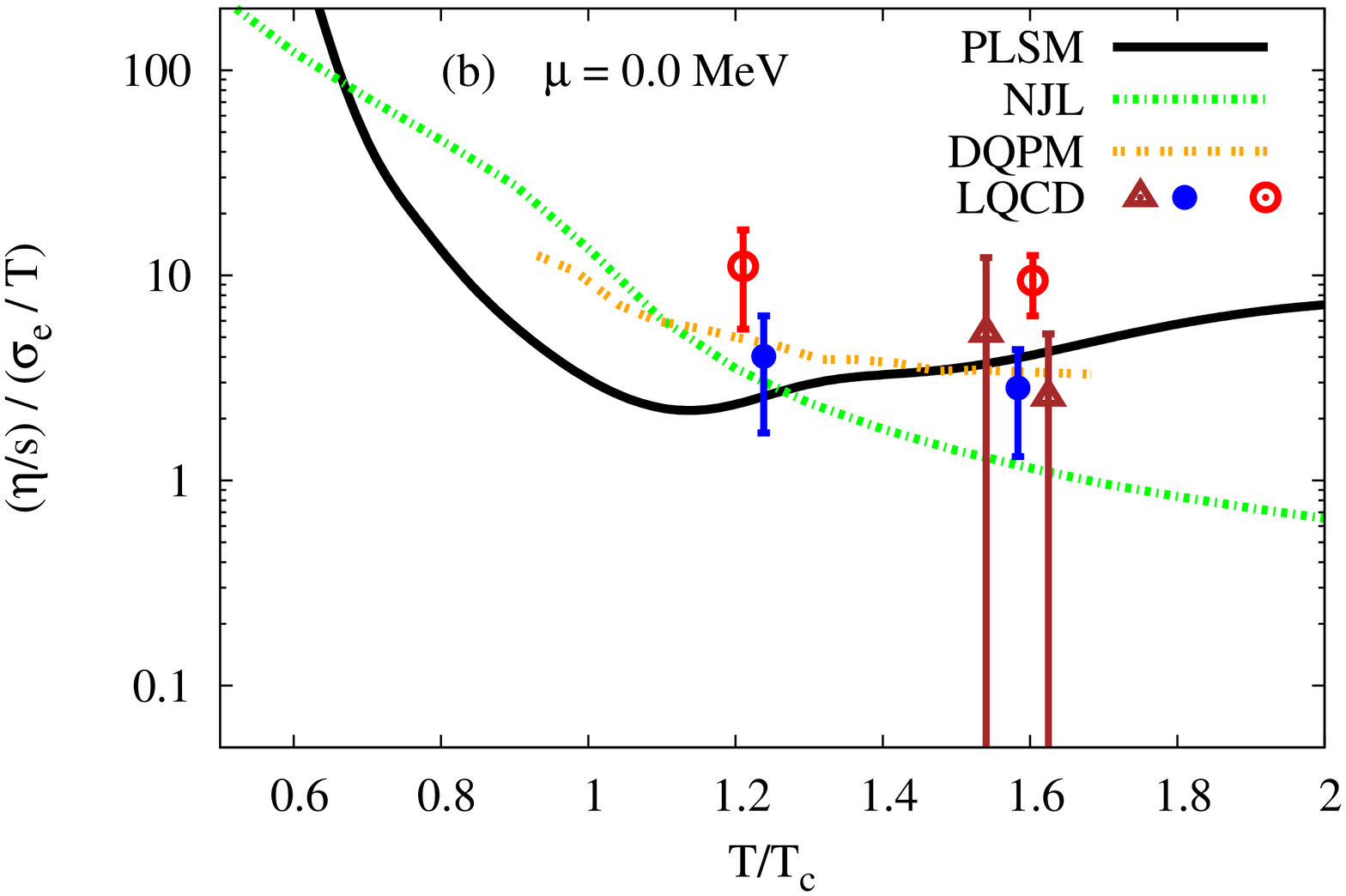}
\caption{\footnotesize In a log-scale, the temperature  dependence of $\xi /s$-to-$\sigma_e /T$  (left-hand panel (a)) and $\eta /s$-to-$\sigma_e /T$ ratios (right-hand panel (b)) is calculated in PLSM approach as functions of temperature at vanishing chemical potential. In left-hand panel, the results are confronted to the available lattice QCD calculations \cite{LQCD:ratioA,LQCD:ratioB,LQCD:ratioC}.  
\label{cond_vis}}}
\end{figure}

In Fig. \ref{cond_vis}, the ratio of bulk and shear viscosity (each normalized to entropy) with the electrical conductivity (each normalized to $T$), $(\xi/s)/ (\sigma_e/T)$ and $(\eta/s)/(\sigma_e/T)$, respectively, is calculated from PLSM as functions of temperature. All these quantities are dimensionless. These predictions probably allow the check of possible scenarios for the QGP formation or when the confinement dynamics derived the hadron into parton matter. Both quantities $(\xi /s)/(\sigma_e/T)$ and $(\eta/s)/(\sigma_e/T)$ can be seen as pivotal insights in understanding the quarks and gluons role in the formation and the evolution of QGP. At $T \gg T_c$, their temperature dependences remain constant \cite{Puglisi:2014}. The latter would be signatures of the unknown properties of QGP \cite{Puglisi:2014}. To the authors' best knowledge,  $(\xi/s)/(\sigma_e/T)$ is not yet calculated in the lattice QCD.

It is found that the temperature dependence of $(\eta/s)/(\sigma_e/T)$ is independent on the strong running coupling \cite{Puglisi:2014}. As the gluons are not charged, this ratio could be regulated by relative strength and chemical composition of the QGP \cite{Puglisi:2014} at very high temperature, i.e., $(5-10)\, T_c$ \cite{Puglisi:2014}. The reported agreement between PLSM results on  $(\eta/s)/(\sigma_e/T)$ and available lattice simulations is very convincing.

\section{Conclusions and outlook}
\label{Conclusion}

The conductivity and viscous properties are essential ingredients for the full characterization of strongly interacting QCD matter. When the QCD system is perturbed from its equilibrium, the transport properties, such as bulk viscosity ($\xi$), shear viscosity ($\eta$), electrical conductivity ($\sigma _e$) and thermal (heat) conductivity ($\kappa$), can be determined. For the sake of completeness, we mention that another transport coefficient that plays an important role in the hydrodynamical evolution of the QCD matter especially around phase transition, is the ratio of bulk viscosity to the thermal entropy ($\eta/s$). 

We have employed SU($3$) PLSM in charactering the temperature dependence of the transport coefficients,  $\xi$, $\eta$, $\sigma _e$, and $\kappa$. In doing this, various thermodynamic quantities such as trace anomaly, speed of sound and specific heat play essential roles. Also, the first- and second-order moments of the quark multiplicities, which are nothing but the quark number multiplicities and their susceptibilities, respectively, contribute to the estimation of the transport coefficients. In determining various PLSM parameters, we have assumed a global minimization of the real potential, Eq. (\ref{potential}). We have evaluated various PLSM order-parameters including chiral condensates and Polyakov-loop fields of light- and strange-quarks. For simplicity, we assume vanishing baryon chemical potential ($\mu=0$). 

The dependence of the relaxation time on the temperature is modelled. We use Eq. (\ref{RlaxTime}), in which the quark number is determined by defining the effective distribution-function in the presence of Polyakov-loop potential. We also use the energy-momentum dispersion-relation, where quark masses are coupled to the sigma fields for light- and strange-quarks, $\sigma_l$ and $\sigma_s$, respectively. We emphasize that the relaxation time is an essential part in our calculations for the  transport coefficients from PLSM, Eqs. (\ref{electric_cond}), (\ref{heatcond}), (\ref{zetaTMU}), and (\ref{etaTMU}). That our PLSM results agree well with first-principle lattice calculations refers to good modelling for the relaxation time and precise estimation for various PLSM parameters. 

With the introduction of the Polyakov-loop potential to LSM, which have two main characteristics, namely the pure gauge potential and the absence of gluon interaction, we have compared the current results with the available lattice calculations. We have found that the SU($3$) PLSM fits well with the lattice QCD calculations, especially from the HOT-QCD collaboration. The chiral subtracted condensates and the order parameters are in good agreement with recent lattice calculations. For instance, they confirm the steeper decrease around the phase transition. 

The electrical conductivity has been estimated for a QCD system, in which the scattering of quarks and gluons are conjectured to be either elastic or inelastic. This allowed the estimation of the relaxation times and decay widths. We have deduced the dimensionless electrical conductivity normalized to $T^2$ and confronted this to recent lattice QCD. We conclude that the present calculations, which combines the quark number multiplicity and their masses, agree well with the lattice QCD simulations, especially above $T_c$, while the DQPM and NJL obviously agree with  below $T_c$.

The temperature dependence of the PLSM heat conductivity normalized to $T^2$ is compared with NJL and DQPM calculations. These different approaches have different critical temperatures, for example, in PLSM $T_c \sim 240~$MeV,  NJL $T_c \sim 200~$MeV, and  DQPM $T_c \sim 158~$MeV. We found that NJL results decrease with the temperature faster than the one from PLSM. From DQPM, the temperature dependence is the opposite to both PLSM and NJL. There are no lattice QCD calculations to compare with them.  Alternatively, we have compared our calculations on the specific heat with recent lattice QCD calculation. An excellent agreement, especially at low and high temperatures, is found. A peak at $T_c$ can be interpreted from the definition of the specific heat;  $c_v =\partial \epsilon/\partial T$, where the rapid change in the energy density ($\epsilon$) around $T_c$ region gives a plausible interpretation of the observed peak in $c_v$. 

The bulk viscosity which is related to various thermodynamic quantities, is calculated from the Green-Kubo correlation. This quantity is strongly related to the phase transition and apparently responses to the instability in the hydrodynamic flow of the resulting partonic plasma. At temperatures close to $T_c$, the ratio of bulk viscosity to entropy rapidly decreases and a first-order phase-transition can be characterized. There is a good agreement with the lattice QCD calculations. The sharp increase, especially when approaching the phase transition, is conjectured to induce instability in the hydrodynamic flow of the partonic plasma. 

The shear viscosity  is assumed as a very suitable tool sensitive to the quark-hadron phase-transition. A good agreement with the lattice QCD calculations and with other effective QCD-like models was observed.  It is noteworthy referring to the low value of the ratio of shear viscosity to entropy ratio. This is caused by the strong interactions and the decrease in the effective degrees of freedom.  The ratio of shear viscosity and the electrical conductivity is confronted to recent lattice QCD calculations. A good agreement can be reported. In particular, the low values of shear viscosity-to-entropy ratio are assumed to reflect low QGP viscosity caused by the stronger interactions and the reduction in the partonic degrees of freedom. 

We argued that the ratio of both bulk and shear viscosities (each is normalized to the thermal entropy) and the dimensionless electrical conductivity, $(\xi/s)/(\sigma_e/T)$ and $(\eta /s)/(\sigma_e/T) $ would be able to favor or disfavor different phenomenological scenarios from PLSM (present work) or PNJL or DQPM, especially when QGP cools down towards $T_c$. Their different independent behaviors of the the different properties of QGP are regulated by the relative strength and the chemical composition of the QGP.  The transport coefficients for these different QCD-like models are apparently depending on the temperature and then chemical potential. 

We conclude that the present PLSM results on various transport properties,  bulk viscosity ($\xi$), shear viscosity ($\eta$), electrical conductivity ($\sigma_e$) and thermal conductivity ($\kappa$), show the essential ingredients that these properties would add to the study of hot and dense QCD matter. Our estimations for various PLSM parameters, chiral condensates, deconfinement order-parameters and the temperature dependence of the relaxation time seem to make the QCD-like approach, PLSM, able to reproduce lattice ''static'' and ''transport'' properties.

\section*{Acknowledgement}

This work is partly supported by the World Laboratory for Cosmology And Particle Physics (WLCAPP), http://wlcapp.net/. AT would like to thank Elena Bratkovskaya and Rudy Marty for providing  PNJL and DQPM results and for the fruitful discussion! The authors gratefully thank the anonymous referee for her/his constructive comments and fruitful suggestions which helped to improve the paper!



\end{document}